\documentclass[reprint,amsmath,amssymb,aps,prb]{revtex4-1}
\usepackage[T1]{fontenc}
\usepackage{graphicx}
\usepackage{mathrsfs}
\usepackage{bm}
\usepackage{hyperref}
\usepackage{dcolumn}
\usepackage{amsmath}
\usepackage{amssymb}
\usepackage{xcolor}
\usepackage{subfig}
\usepackage{caption}
\usepackage{siunitx}
\usepackage{ulem}
\usepackage{textcomp}
\hypersetup{colorlinks=true, linkcolor=blue, citecolor=red, urlcolor=blue}

\everymath{\displaystyle}

\newcommand{\gggg}{\pmb{g}}
\newcommand{\KK}{\pmb{K}}
\newcommand{\kk}{\pmb{k}}
\newcommand{\qq}{\pmb{q}}
\newcommand{\dd}{\pmb{d}}
\newcommand{\rr}{\pmb{r}}

\newcommand{\AAA}{\pmb{A}}
\newcommand{\aaa}{\pmb{a}}
\newcommand{\ssigma}{\pmb{\sigma}}
\newcommand{\mC}{\mathcal{C}}

\newcommand{\mR}{\mathcal{R}}
\newcommand{\GG}{{\rm GG}}
\newcommand{\GBN}{{\rm GBN}}
\newcommand{\Ket}[1]{\left|{#1}\right\rangle}
\newcommand{\Bra}[1]{\left\langle{#1}\right|}

\begin{document}

\title{Moir\'e Commensurability and the Quantum Anomalous Hall Effect \\ in Twisted Bilayer Graphene on Hexagonal Boron Nitride}
\author{Jingtian Shi}
\affiliation{Department of Physics, University of Texas at Austin, Austin TX 78712, USA}
\author{Jihang Zhu}
\affiliation{Department of Physics, University of Texas at Austin, Austin TX 78712, USA}
\author{A.H. MacDonald}
\affiliation{Department of Physics, University of Texas at Austin, Austin TX 78712, USA}

\date{\today}

\begin{abstract}
    The quantum anomalous Hall (QAH) effect is sometimes observed in twisted bilayer graphene (tBG) when it is nearly aligned with an
    encapsulating hexagonal boron nitride (hBN) layer.  We propose that the appearance or absence of the QAH effect in individual
    devices could be related to commensurability between the graphene/graphene and graphene/hBN moir\'e patterns.
    We identify a series of points in the $(\theta_{\GG},\theta_{\GBN})$ twist-angle
    space at which the two moir\'e patterns are commensurate, allowing moir\'e band theory to be applied, and show that
    the band Chern numbers are in this case sensitive to a rigid in-plane hBN displacement.  Given this property,
    we argue that the QAH effect is likely only when i) the $(\theta_{\GG},\theta_{\GBN})$ twist-angle-pair is close enough to a
    commensurate point that the two moir\'e patterns yield a supermoir\'e pattern with a sufficiently long length scale,
    and ii) the supermoir\'e has a percolating topologically non-trivial QAH phase.
    For twist angles far from commensurability, the hBN layer acts as a source of disorder
    that can destroy the QAH effect.  Our proposal can explain a number of current experimental observations.  Further experimental
    studies that can test this proposal more directly are suggested.
\end{abstract}

\maketitle

\section{Introduction}

Two graphene sheets that have a small orientational misalignment (twisted bilayer graphene - tBG)
form a quasiperiodic moir\'e superlattice, whose electronic structure is well-described by moir\'e band theory. \cite{Rafi2011} Correlated insulating states, \cite{Cao2018Correlated}
Chern insulators, \cite{Sharpe2019,Serlin2020,Kevin2020,Saito2020} and
superconductivity \cite{Cao2018_SC,Lu2019,Yankowitz2019,Saito2020_independent} have been observed in tBG when the twist angle is close to a
magic angle that enables strong correlation physics associated with exceptionally flat moir\'e bands.
The introduction of twist angle as a new tunable degree of freedom
has now been exploited to create strong correlations in a variety of different
multi-layer van der Waals systems. \cite{Chen2019_SC,Shen2020,Cao2020_TBBG,Spanton2018,LeiWang2020,Regan2020,Liu2020}

Recent experiments have shown both non-quantized \cite{Sharpe2019} and quantized \cite{Serlin2020,Tschirhart2020} anomalous Hall effects
can occur in magic angle twisted bilayer graphene when at least one graphene layer is nearly aligned with an
encapsulating hexagonal boron nitride (hBN) layer, and the number of carriers per moir\'e period is close to an odd integer.
The anomalous Hall effect is normally understood in terms of a mean-field picture, in which it arises
from a combination of spontaneous valley polarization and non-zero Chern numbers of the valley-projected flat moir\'e bands induced by violation of inversion symmetry ($\mC_2$).\cite{Ming2018,HoiChun2018,Yahui2019_flatChern,YahuiZhang2019,Nick2020}
Somewhat mysteriously, the anomalous Hall effect is not always present even with hBN alignment.

The theoretical description of hBN encapsulated tBG runs into a fundamental difficulty when one or both
hBN layers are nearly aligned with the tBG layers.  Because of
the small lattice constant mismatch between graphene and hBN, the nearly-aligned hBN layers produce
additional moir\'e patterns \cite{Moon2014,Wallbank2015,Jeil2014,Jeil2015,Jeil2017,Lin2019} which are
not in general commensurate with the moir\'e pattern of tBG.
Therefore, the low-energy Hamiltonian is only quasi-periodic, disallowing all the
simplifications that come from Bloch's theorem.
Similar moir\'e pattern interplays can also arise in twisted trilayer graphene.\cite{ZiyanZhu2020_tTG}
Most of the existing theoretical work on the anomalous Hall effect \cite{YahuiZhang2019,Nick2020} and related properties \cite{Fengcheng2020,Cecile2020,Jihang2020_Magnetic,Yahya2019,Shubhayu2020,ChengPing2020} of tBG/hBN and hBN/tBG/hBN systems
employs a highly simplified model in which only the spatially average sublattice energy difference is retained in the graphene/hBN moir\'e potentials. The justification for this expediency is not obvious, since the spatially averaged and position-dependent
tBG/hBN interaction terms have similar energy scales \cite{Jeil2017} and are therefore at first sight equally important.

The aim of this paper is to study the effect of the interplay between the moir\'e patterns on the anomalous Hall effect of
encapsulated tBG.  For definiteness we will assume that only one of the encapsulating hBN layers is aligned, which allows
us to restrict our attention to tBG/hBN trilayers.  In mean-field theory spontaneous valley polarization
occurs when the moir\'e bands are sufficiently narrow to satisfy a Stoner criterion.  It follows that both
criteria for a quantized anomalous Hall effect, topologically non-trivial valley-projected bands and valley polarization,
are simply related to the electronic structure issues on which we focus.

We notice that at particular combinations of the two twist angles, $\theta_{\GG}$ between the two graphene layers
and $\theta_{\GBN}$ between the hBN  and its adjacent graphene layer, the two moir\'e patterns are commensurate.  The system is then
periodic in a larger unit cell, allowing the use of Bloch's theorem with both moir\'e patterns present.
Recent papers \cite{Cea2020,Lin2020} have noticed several such commensurate points in the twist angle space
and we provide a general description of all commensurate geometries.
For a commensurate system, rigid translation of the hBN layer by $\dd$ at fixed twist angle changes the moir\'e band structures,
and even moir\'e band Chern numbers.\cite{Cea2020}  We characterize this dependence in terms of maps of Chern numbers and bandwidths
\textit{vs.} $\dd$, from which electronic properties can be estimated.

A \textit{supermoir\'e} pattern, also known as a \textit{moir\'e of moir\'e}, is formed when the two moir\'e patterns are nearly,
but not exactly commensurate. Supermoir\'e electronic structure has been studied in hBN/graphene/hBN trilayers \cite{WangZihao2019,Nicolas2019,Andelkovic2020} and in
twisted trilayer graphene, \cite{ZiyanZhu2020_relax,ZiyanZhu2020_tTG,KanTing2019} but not yet in tBG/hBN.
We point out here that the supermoir\'e can be viewed as a commensurate structure with spatially varying $\dd$. Thus its electronic properties can be well described by a local moir\'e band picture, where local properties are defined by the local Hamiltonian $H(\rr)=H(\dd(\rr))$, with $H(\dd)$ the Hamiltonian of the commensurate structure.  In this picture, the Chern number \textit{vs.} $\dd$ map expands to a spatial Chern number
phase pattern, which is reminiscent of the percolation\cite{Trugman1983} picture and of the
Chalker-Coddington model \cite{Chalker_1988,Marston1999} of the quantum Hall effect.
In the present case, however, there are also semimetal phases due to overlaps between the valence and conduction bands that are
indirect in momentum space.  For the quantum anomalous Hall (QAH) effect
the possible presence of regions in which the Stoner criterion for spontaneous valley polarization
is not satisfied because of locally larger bandwidths is also relevant.

In this local picture a global QAH effect can appear only if the following two conditions
are satisfied: i) the supermoir\'e period must be long enough that edge states between
topologically distinct phases do not couple to each other and ii) a topologically nontrivial insulating phase must
percolate across the device.  The first condition is always satisfied over a finite range of twist angles
close to a commensurate point and the second condition can usually be satisfied by varying the electrical
potential difference $U$ between layers by applying a gate-controlled out-of-plane electric field.

In the opposite limit in which the two moir\'e patterns are far from being commensurate and
the local moir\'e band picture fails, we assume that the moir\'e periodic part of the hBN potential acts
like a disorder potential.  The moir\'e bands of tBG are then widened by scattering from the hBN potential.
In some cases this broadening effect may also make the full bandwidth exceed the interaction strength,
standing in the way of spontaneous valley polarization and therefore of the anomalous Hall effect.
Our proposals provide a possible explanation for a number of experimental observations,
but are not conclusively established by exisiting experiments.

This paper is organized as follows: In Sec. \ref{sec_geometry}
we first identify the commensurate twist angle pairs, and then discuss the geometry of tBG/hBN supermoir\'e systems
in terms of proximity to these commensurate points.
In Sec. \ref{sec_cmsrHam} we describe the continuum model we use to investigate the electronic structure.
In Sec. \ref{sec_cmsrAH} we present our results for the spatial pattern of tBG/hBN supermoir\'e's phases
calculated from our model in a local-band approximation.
In Sec. \ref{sec_twistanglewindows} we estimate the twist angle windows within which
QAH effects can occur in tBG/hBN supermoir\'e.
In Sec. \ref{sec_incmsr} we analyze the limit in which the two moir\'e patterns are far from being
commensurate.  Then in Sec. \ref{sec_discuss},
we use our results to provide possible explanations of current experiments and suggest further experimental approaches to test our proposals in the future.
Sec. \ref{sec_summary} contains the summary and main conclusions of this paper.

\begin{figure*}
    \centering
    \includegraphics[width=0.96\textwidth]{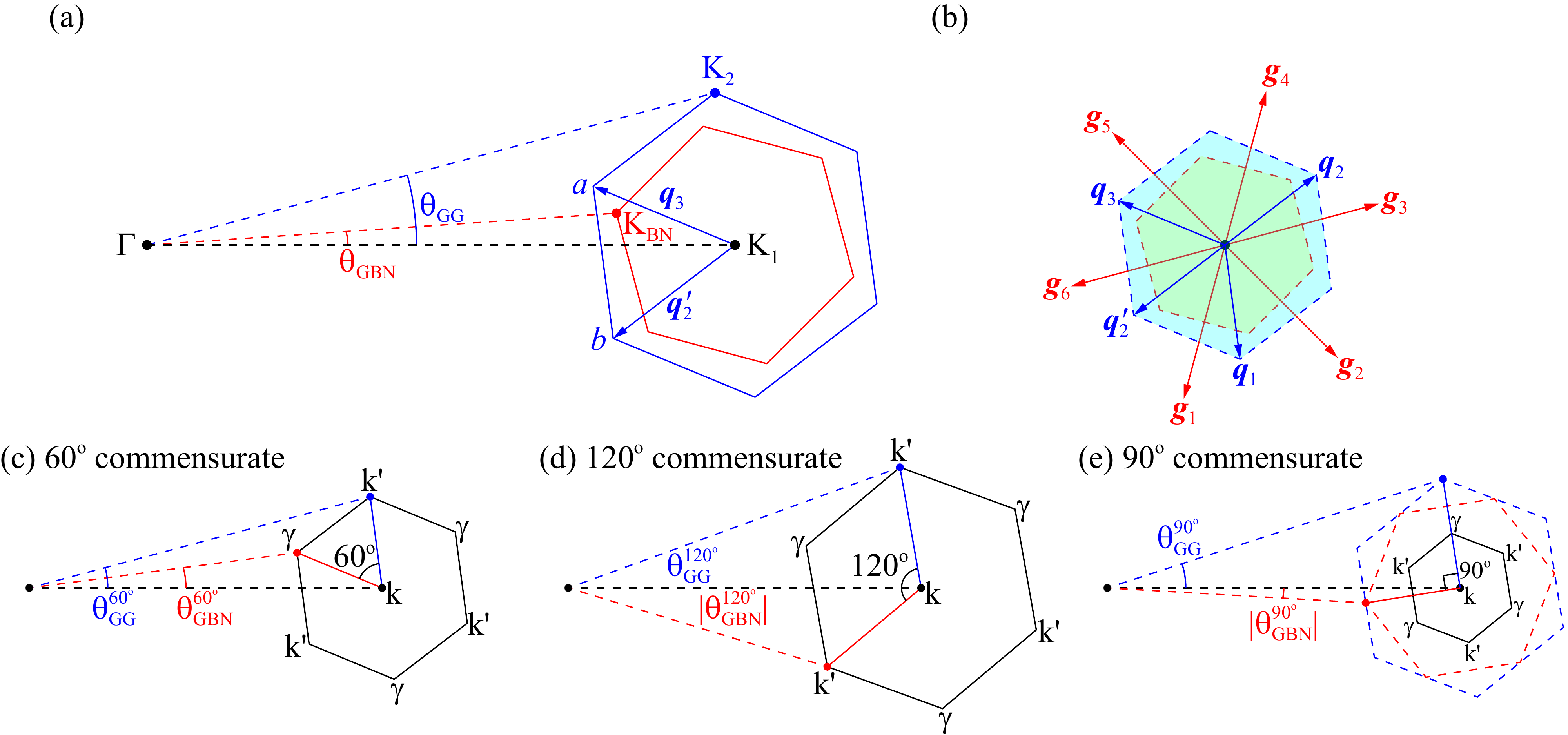}
    \caption{Schematic reciprocal space geometry of a tBG/hBN system:
    (a)-(b) Generic twist angles: $K_1$, $K_2$ and $K_{\rm BN}$ are Brillouin-zone corner points of
    graphene layer 1, graphene layer 2, and hBN, respectively.  The mBZs of the
    G1/G2 and G1/hBN moir\'e patterns are illustrated by blue and red hexagons respectively;
    (c) $60^\circ$ commensurate, (d) $120^\circ$ commensurate, and (e) $90^\circ$ commensurate systems.  In (e), the blue dashed, red dashed, and inner black solid hexagons are respectively the mBZs of the G1/G2, G1/hBN heterojunctions and the entire trilayer.
    The high-symmetry points of the mBZs of the commensurate systems are given their conventional labels.}
    \label{fig_geometry}
\end{figure*}

\section{Geometry}
\label{sec_geometry}

\subsection{Commensurate tBG/hBN}

We consider a tBG/hBN trilayer system in which
the graphene layer adjacent to the nearly-aligned hBN layer is labeled as layer 1 or G1, while the top graphene layer is
labeled as layer 2 or G2.
We let G2 and the hBN layer both be twisted
relative to G1 by small angles, denoted respectively as $\theta_{\GG}$ and $\theta_{\GBN}$.
The lattice constant of microscopic graphene honeycomb $a_{\rm G}$ is
taken to be $a_{\rm G}=\sqrt{3}\times1.42\text{\AA}$,\cite{Castro2009_grapheneRMP}
$\alpha=a_{\rm BN}/a_{\rm G}=1.017$ \cite{LiuLei2003_hBN} is the ratio between the
hBN and graphene lattice constants, and the $A$ sublattice of hBN is taken to be occupied by boron atoms.

The moir\'e patterns of the G1/G2 and G1/hBN heterojunctions are commensurate if and only if their moir\'e
reciprocal lattices are commensurate.  We show in Appendix \ref{sec_app_geometry} that the commensurabilty condition is:
\begin{equation}
    n(\KK_{\rm BN}-\KK_1)=p\qq_3+q\qq_2'
    \label{eq_cmsr_condition},
\end{equation}
where $(n,p,q)$ is a triplet of coprime integers that
characterizes distinct commensurate structures.  Here $\KK_1=(4\pi/3a_{\rm G},0)$ and $\KK_{\rm BN}=(4\pi/3a_{\rm BN})(\cos\theta_{\GBN},\sin\theta_{\GBN})$ are the Dirac points of graphene layer 1 and hBN respectively, and $\qq_3$ and $\qq_2'$ are defined in Figs. \ref{fig_geometry} (a)-(b).
For given $n$, $p$ and $q$, the twist angle pair $(\theta_{\GG},\theta_{\GBN})$ is implied by Eq. (\ref{eq_cmsr_condition}).
In the small twist angle approximation $(\cos\theta,\sin\theta)\rightarrow(1,\theta)$,
\begin{equation}
    \theta_{\GG}\approx\frac{n}{p+q}\times1.1^\circ, \quad \theta_{\GBN}\approx\frac{p-q}{p+q}\times0.55^\circ .
\end{equation}
Exact expressions for the commensurate twist-angle pairs are discussed in Appendix \ref{sec_app_geometry}.

We focus our attention on integer triplets that satisfy $n=p+q$.  Geometrically these triplets
correspond to the case in which $\KK_{\rm BN}$ is on the line $\overleftrightarrow{ab}$ illustrated in Fig. \ref{fig_geometry} (a).
We choose these commensurate structures because they
yield a tBG angle $\theta_{\GG}\approx1.1^\circ$ that is very close to the magic angle $\sim1.05^\circ$. \cite{Rafi2011}
Only for these twist angles do we expect the strong correlation physics  \cite{Cao2018_SC} that is responsible for
much of the interest in tBG/hBN systems to appear.  For this series of
commensurate points, the area of the supercell
is $N=n^2$ times larger than the corresponding tBG system.  It follows that
each moir\'e band of isolated tBG is split into $n^2$ bands by coupling to the adjacent hBN layer.
We note that commensurate points are dense in twist angle space, just as rational numbers are dense on the real line.
However, most commensurate points have very large $n$, which means that the tBG bands are split into a correspondingly
large number of subbands, and are therefore unlikely to lead to observable consequences in finite-size systems with non-zero disorder.
We therefore focus on the discrete set of low-order commensurate points that we have identified.
Two different $n=1$ systems have been identified in previous work: $(p,q)=(1,0)$ \cite{Cea2020} and
$(p,q)=(2,-1)$.\cite{Lin2020} Figures \ref{fig_geometry} (c)-(e) show schematics of several of the simplest structures in this series,
which we will refer to respectively as $60^\circ$ commensurate (Fig. \ref{fig_geometry} (c), $(n,p,q)=(1,1,0)$, $\theta_{\GBN}\approx0.55^\circ$),
$120^\circ$ commensurate (Fig. \ref{fig_geometry} (d), $(n,p,q)=(1,0,1)$, $\theta_{\GBN}\approx-0.55^\circ$),
and $90^\circ$ commensurate (Fig. \ref{fig_geometry} (e), $(n,p,q)=(2,1,1)$, $\theta_{\GBN}\approx0^\circ$).

\begin{figure*}
    \centering
    \includegraphics[width=0.96\textwidth]{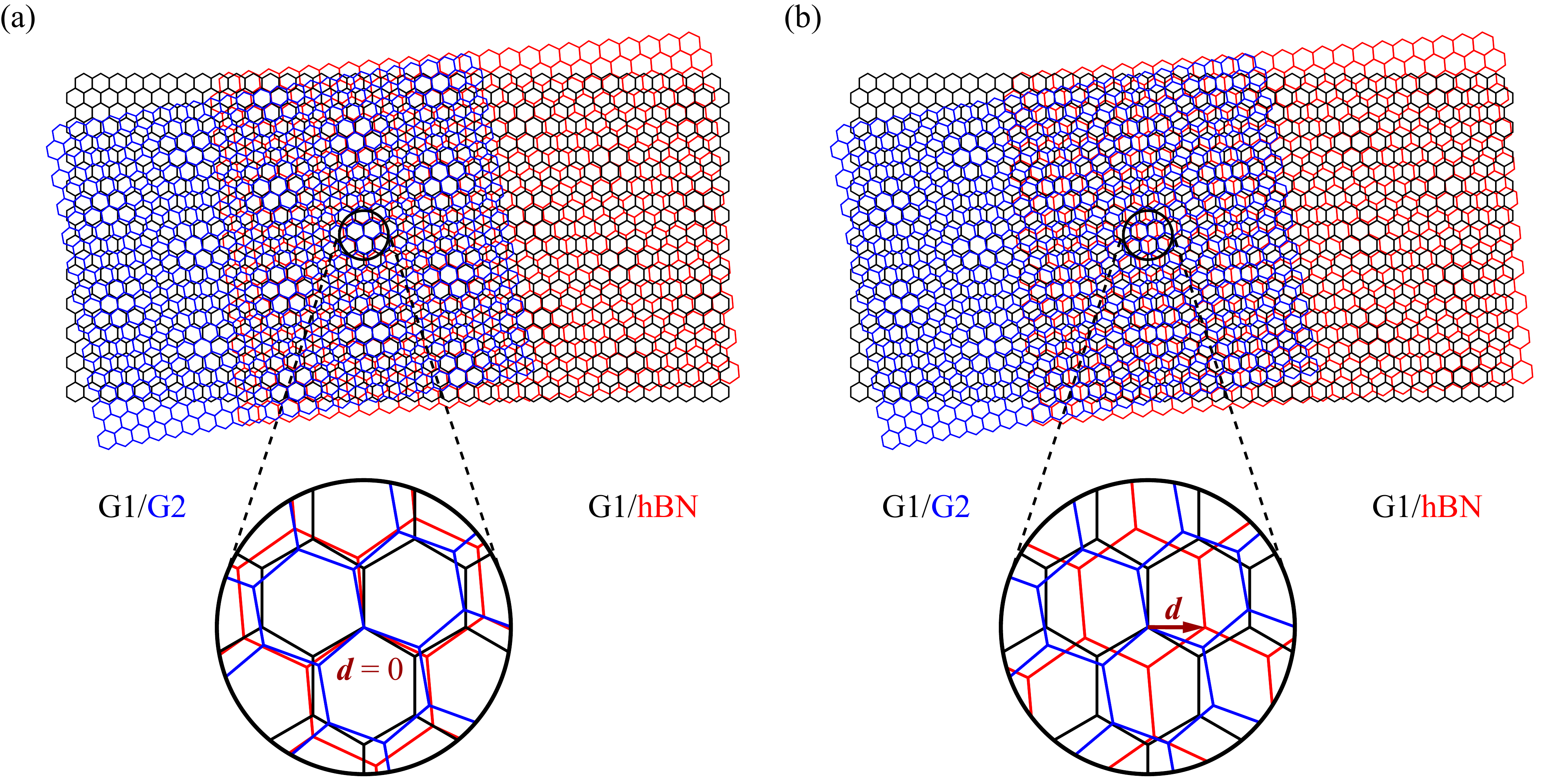}
    \caption{Schematic illustration of two moir\'e patterns that differ by a rigid displacement $\dd$ of the hBN layer of a
    commensurate tBG/hBN system: (a) $\dd=0$; (b) $\dd\neq0$. As we see, the G1/hBN moir\'e pattern is shifted.
    The twist angles and lattice constant mismatches are exaggerated in this schematic.}
    \label{fig_stacking}
\end{figure*}

In commensurate tBG/hBN trilayers, electronic properties change when one moir\'e pattern is laterally
translated relative to the other by a rigid in-plane translation of any one of the three layers.
This contrasts with the bilayer moir\'e superlattice case in which the effect of translating one of the two layers is simply to produce a
magnified global shift of the moir\'e pattern, which has no consequence in the thermodynamic limit.
In a trilayer, shifting an outside layer only shifts one of the two moir\'e patterns, and
shifting the middle layer shifts both, but not necessarily by the same amount.
In this paper, we fix a local AA stacking point of the G1/G2 moir\'e pattern at the origin and
examine how electronic structure changes when the
hBN layer is translated by $\dd$ relative to a point at which
its $A$ (boron) site is at the origin (see Fig. \ref{fig_stacking}).
A shift in the hBN layer by $\dd$ shifts the G1/hBN moir\'e pattern by
\begin{equation}
    \dd_M=\left(1-\alpha\mR_{\theta_{\GBN}}\right)^{-1}\dd.
    \label{eq_dM_d}
\end{equation}
(Here $\mR_\theta$ is an operator that rotates a vector counterclockwise by $\theta$.)

\subsection{tBG/hBN supermoir\'e structures}

A supermoir\'e structure is formed when the two twist angles are displaced slightly away from a
low-order commensurate point, \textit{i.e.} when
\begin{equation}
    \theta_{\GG}=\theta_{\GG}^{npq}+\delta\theta_{\GG}, \quad\theta_{\GBN}=\theta_{\GBN}^{npq}+\delta\theta_{\GBN},
\end{equation}
where $(\theta_{\GG}^{npq},\theta_{\GBN}^{npq})$ is the commensurate pair defined by the integer triplet $(n,p,q)$ defined in
Eq. (\ref{eq_cmsr_condition}), and both $\delta\theta_{\GG}$ and $\delta\theta_{\GBN}$ are $\sim0.01^\circ$.
The period and orientation of the supermoir\'e pattern depend on both $\delta\theta_{\GG}$ and $\delta\theta_{\GBN}$.

For sufficiently large supermoir\'e periods, the supermoir\'e structure can be characterized in terms of local commensurate
tBG/hBN systems with the shift parameter $\dd$ varying slowly in space.
We let $\dd=0$ correspond to local AAA stacking at $\rr=0$, since in the supermoir\'e case
a global shift of the hBN layer $\dd(\rr)\rightarrow\dd(\rr)+\dd_0$ does not affect the overall supermoir\'e pattern.
This can be seen by noting that a shift of hBN causes a magnified shift of the G1/hBN moir\'e pattern, which in turn
produces a further magnified shift of the supermoir\'e pattern, and this can be cancelled by a reselection of the origin.

The analysis in Appendix \ref{sec_app_geometry_sm} shows that in the small twist angle limit
the magnification factor from $\dd$ to $\rr$ is
\begin{equation}
    \gamma\equiv\frac{|\rr|}{|\dd|}\approx\frac{n}{\left|n\delta\theta_{\GBN}-\left(pe^{i\pi/3}+qe^{2i\pi/3}\right)\delta\theta_{\GG}\right|},
    \label{eq_gamma}
\end{equation}
and that when the supercell of the $(n,p,q)$ commensurate system contains $N$ moir\'e cells of tBG,
the ratio $r_a$ between the supermoir\'e lattice constant $a_{\rm sm}$ and the hBN lattice constant $a_{\rm BN}$ is
\begin{equation}
    r_a\equiv\frac{a_{\rm sm}}{a_{\rm BN}}=\frac{\gamma}{\sqrt{N}}.
    \label{eq_ra}
\end{equation}
For supermoir\'es near $60^\circ$, $120^\circ$ and $90^\circ$ commensurate points,
\begin{equation}
    r_a^{120^\circ,60^\circ}=\frac{1}{\sqrt{\delta\theta_{\GG}^2+\delta\theta_{\GBN}^2\pm\delta\theta_{\GG}\delta\theta_{\GBN}}}
\end{equation}
with the $+$ sign for $120^\circ$, and
\begin{equation}
    r_a^{90^\circ}=\frac{1}{\sqrt{3\delta\theta_{\GG}^2+4\delta\theta_{\GBN}^2}}.
\end{equation}

\section{Electronic Properties}
\label{sec_elecprop}

\subsection{Model Hamiltonian}
\label{sec_cmsrHam}

In this section we describe how we model tBG/hBN trilayers with arbitrary twist angles $\theta_{\GG}$ and $\theta_{\GBN}$ and hBN layer
translations $\dd$. We adopt the commonly employed non-interacting model Hamiltonian, focusing on one valley since the other valley can be easily obtained by time reversal.  The low-energy degrees of freedom are entirely in the graphene bilayer, but have a periodic contribution due to the adjacent hBN layer that we separate by writing
\begin{equation}
    H(\dd)=H_{\rm tBG}+V_{\rm BN}(\dd).
    \label{eq_hdd}
\end{equation}
The bilayer has four $\pi$-electron sublattices counting the two honeycomb layers.
For $H_{\rm tBG}$ we use the well-known four-sublattice continuum model Hamiltonian of tBG, \cite{Rafi2011}
adding a gate-controlled interlayer potential difference $U$.
We adopt the \textit{ab initio} estimates for the same and different sublattice interlayer tunneling parameters in tBG by
setting $w_{AB}=113\rm meV$ \cite{Jeil2014} and $w_{AA}/w_{AB}=0.8$, a value that
accounts approximately for lattice relaxation. \cite{Stephen2019}

In Eq. (\ref{eq_hdd}) we assume that $V_{\rm BN}(\dd)$ is non-zero only on G1 layer and not on G2.
$V_{\rm BN}(\dd)$ can be separated\cite{Jeil2014} into a spatially averaged term that is independent of position, and
a periodic contribution:
\begin{equation}
    V_{\rm BN}(\dd)=\sum_{\kk}\left(\psi_{1\kk}^\dag(m_0\sigma^z)\psi_{1\kk}+\sum_{j=1}^6\psi_{1\kk}^\dag V_j(\dd)\psi_{1(\kk+\gggg_j)}\right).
    \label{eq_VBN}
\end{equation}
The first term on the right hand side (RHS) of Eq. (\ref{eq_VBN}) captures the critical broken inversion symmetry in
the G1 layer, as discussed in previous work. \cite{YahuiZhang2019,Nick2020,Fengcheng2020,Cecile2020,Jihang2020_Magnetic,Yahya2019,Shubhayu2020,ChengPing2020}
\textit{Ab initio} calculations of monolayer graphene/hBN with full lattice relaxation
yield the estimate $m_0=3.62\rm meV$, \cite{Jeil2017} but experiments suggest that
$m_0$ is significantly larger, \cite{Hakseong2018} possibly as large as
$\sim15\rm meV$ \cite{Hunt2013} and possibly reflecting many-body physics that is absent in the
DFT calculation.\cite{Justin2013} Since it is unclear whether many-body enhancement of $m_0$ is also important in tBG/hBN,
we take $m_0=3.62\rm meV$ in most of our explicit calculations, using the value $m_0=10\rm meV$
in some calculations for comparison purposes.

\begin{figure*}
    \centering
    \includegraphics[width=0.96\textwidth]{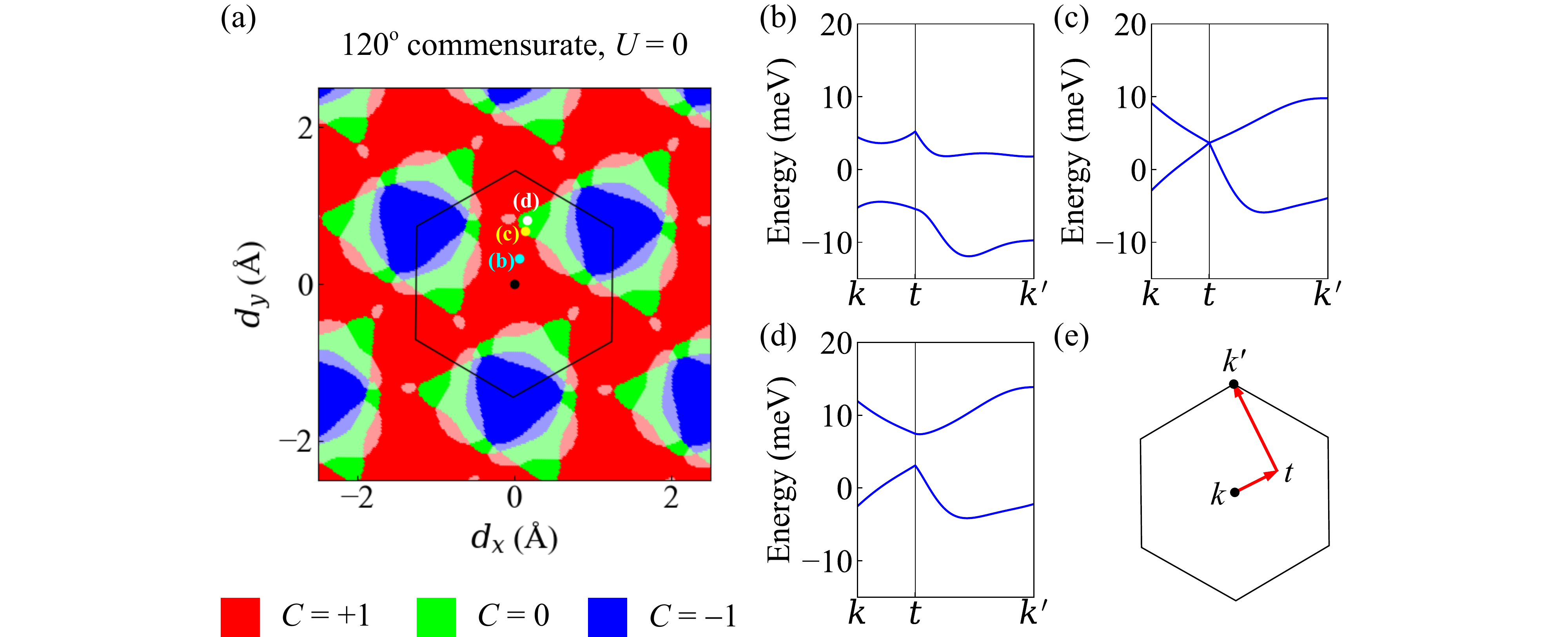}
    \caption{(a) Map of valence band Chern number $C$ \textit{vs.} hBN displacement $\dd$ for a $120^\circ$ commensurate tBG/hBN moir\'e superlattice with zero interlayer potential difference $U$.  Different colors specify different Chern numbers, as
    ilustrated by the legend below. The lighter shades identify semimetal regions with a gap closing that is indirect in momentum space.
    The black hexagon is the Wigner-Seitz cell of the hBN layer.
    (b)-(d) Band structures of the system at the $\dd$ values marked by cyan (b), yellow (c) and white (d) dots in map (a).
    The band structures are plotted along the red path shown in (e), which includes the point $t$ at which the band touching
    occurs in (c).  Band touching always occurs at some point in the mBZ along the map's Chern number region boundaries.}
    \label{fig_C(d)}
\end{figure*}

The second term on the RHS of Eq. (\ref{eq_VBN}) accounts for
the G1/hBN moir\'e pattern. The 6 transfer momenta $\gggg_j$ are from the first shell of the moir\'e reciprocal lattices and the $V_j$'s are matrices that act on sublattice degrees of freedom. \textit{Ab initio} calculation \cite{Jeil2017}
estimate that all $V_j$'s are $\sim 10\rm meV$.  These matrices are detailed in
Appendix \ref{sec_app_Hamiltonian}.
We capture the $\dd$ dependence of the hopping matrix $V_j$ by multiplying the Fourier expansion coefficients by phase factors:
\begin{equation}
    V_j(\dd)=V_j(0)e^{i\gggg_j\cdot\dd_M}
    \label{eq_Vj_d_dependence},
\end{equation}
where the shift $\dd_M$ of the G1/hBN moir\'e pattern depends on $\dd$ via Eq. (\ref{eq_dM_d}).

\subsection{Anomalous Hall effect at commensurate twist-angle pairs and supermoir\'e}
\label{sec_cmsrAH}

Figure \ref{fig_C(d)} (a) contains a map of the valence band Chern number $C$ \textit{vs.} $\dd$ for
the $120^\circ$ commensurate tBG/hBN trilayer implied by the model Hamiltonian described above with $U=0$.
The Chern numbers were calculated using the highly efficient method described in Ref. \onlinecite{Takahiro2005}.
The structure present in the Chern number map demonstrates that band crossings occur as $\dd$ is varied.
In Figs. \ref{fig_C(d)} (b)-(d) we plot the band structures at the $\dd$ points highlighted in
Fig. \ref{fig_C(d)} (a).  The expected band inversion at the Chern number boundary is apparent in these figures.
We emphasize that if the $\gggg_j\ne0$ terms in Eq. (\ref{eq_VBN}) were neglected,
then the spectrum would be independent of $\dd$, and the Chern number map would be monochromatic.  The interesting structure is present only because the
G1/hBN moir\'e pattern has a qualitative influence on electronic structures.

\begin{figure*}
    \centering
    \includegraphics[width=0.96\textwidth]{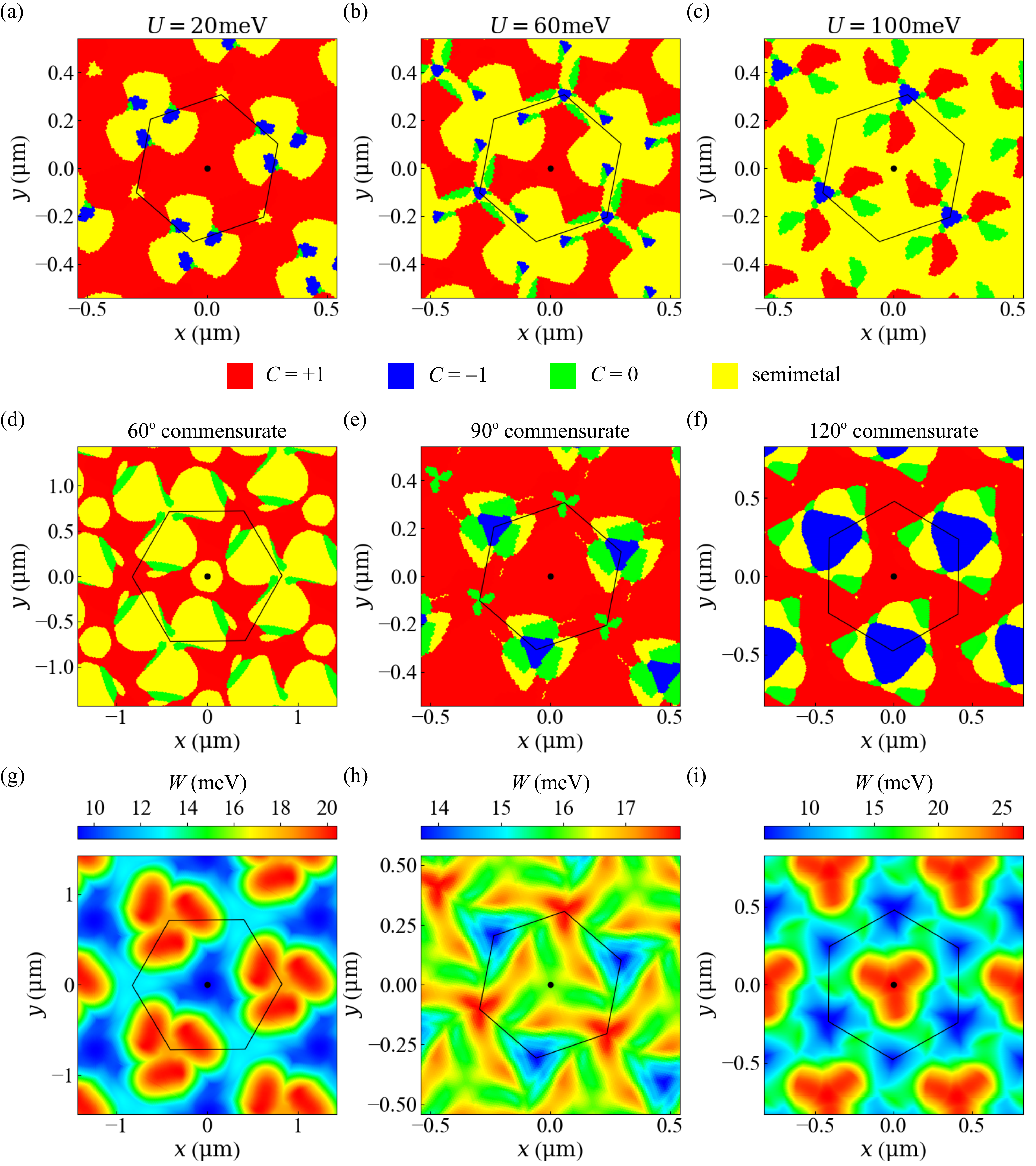}
    \caption{(a)-(c) Phase maps of a supermoir\'e structure close to the $90^\circ$ commensurate twist point with $\delta\theta_{\GG}=\delta\theta_{\GBN}=0.01^\circ$, under various interlayer potential differences $U$.
    Different colors specify different phases, as illustrated by the legend on the bottom.
    $C$ is the valence band Chern number. The black hexagon is the Wigner-Seitz cell of the supermoir\'e pattern.
    At $U=20\rm meV$ the $C=+1$ phase is globally connected, indicating an overall measurable QAH effect.
    Otherwise the quantum Hall conductance is not quantized and the longitudinal conductivity
    is non-zero. (d)-(f) Phase maps of a supermoir\'e structure with
    $\delta\theta_{\GG}=\delta\theta_{\GBN}=0.01^\circ$ near (d) $60^\circ$ commensurate;
    (e) $90^\circ$ commensurate; (f) $120^\circ$ commensurate structures, with $U=-20\rm meV$.
    All three cases have percolating $C=1$ phases. (g)-(i)
    Maps of the local conduction band width $W$ of the same systems as in (d)-(f).
    For $90^\circ$ commensurate, $W$ refers to the difference between the top of the highest miniband and the bottom of the lowest miniband split from the conduction band of tBG.  In a Stoner approximation spontaneous valley polarization occurs when an exchange interaction
    parameter exceeds $W$.}
    \label{fig_phasemaps}
\end{figure*}

When the electronic structure of a supermoir\'e system is described in a local band picture,
the $C(\dd)$ map in Fig. \ref{fig_C(d)} (a) expands to a spatial map $C(\rr)=C(\dd(\rr))$ with magnification
factor $\gamma$ defined in Eq. (\ref{eq_gamma}).
When narrow bands lead to spontaneous valley polarization at odd moir\'e band fillings, \cite{Nomura2006,Ming2018,Jihang2020_Magnetic,Saito2020}
spatial regions with different valley-dependent Chern numbers
will have topologically distinct QAH or trivial phases.
We notice that at some $\dd$'s the valence and conduction bands overlap, giving rise to semimetal regions
that cannot support a quantized Hall conductance, but can in principle support spontaneous valley
polarization and therefore non-zero Hall effects.   The entire supermoir\'e structure is therefore
expected to support a complex spatially inhomogenous state containing
alternating Chern insulator, trivial insulator, and semimetal phases.
Several samples of such patterns are plotted in Fig. \ref{fig_phasemaps} (a)-(f).
We see that at certain interlayer potential differences $U$, the $C=1$ phase or the semimetal phase
percolates, while at other $U$'s no phase percolates. The percolation properties
of different $U$'s are summarized in Table \ref{table_percolation}, where we see that percolation of the $C=1$
phase is most common in nearly $120^\circ$ commensurate systems. If many-body effects do enhance $m_0$ or
the single-particle sublattice splitting term in the Hamiltonian is larger than the
estimate employed for these plots, more $C=1$ percolation is expected
because the original gap opened by the $m_0$ term of the hBN potential is then larger and less easily inverted
by either $U$ or the $\gggg \ne 0$ terms of the G1/hBN moir\'e potential. This
observation is quantified in Appendix \ref{sec_app_largerm0} where the corresponding
results for $m_0=10\rm meV$ are summarized.

\begin{table*}
    \caption{Summary of percolating supermoir\'e phases of different commensurate structures under various interlayer potential difference $U$.
    S labels percolating semimetal states; X labels states with no percolating phase.}
    \begin{tabular}{|c|c|c|c|c|c|c|c|c|c|c|c|}
        \hline $U$ (meV) & $-100$ & $-80$ & $-60$ & $-40$ & $-20$ & 0 & 20 & 40 & 60 & 80 & 100 \\ \hline
        $60^\circ$ commensurate & S & S & $C=1$ & $C=1$ & $C=1$ & X & X & S & S & S & S \\ \hline
        $90^\circ$ commensurate & S & X & X & X & $C=1$ & $C=1$ & $C=1$ & X & X & X & S \\ \hline
        $120^\circ$ commensurate & $C=1$ & $C=1$ & $C=1$ & $C=1$ & $C=1$ & $C=1$ & $C=1$ & X & X & $C=1$ & $C=1$ \\ \hline
    \end{tabular}
    \label{table_percolation}
\end{table*}

So far we have assumed full valley polarization.  In practice valley polarization occurs only if the bands are sufficiently narrow
relative to interaction strength.  In Figs. \ref{fig_phasemaps} (g)-(i) we map the conduction band width $W$ \textit{vs.} position $\rr$.
It follows from the Stoner mean-field
criterion that spontaneous valley polarization is likely to be absent when the bandwidth $W$ exceeds the relevant exchange energy $X$.
Self-consistent Hartree-Fock calculations in previous work suggest that
$X\approx30\rm meV$ in tBG with twist angle $\theta_{\GG}=1.1^\circ$ at moir\'e band filling factor $\nu=1$. \cite{Ming2018}
Since Hartree-Fock calculations tends to overestimate the exchange energy, our $W(\rr)$ maps may
imply that some valley unpolarized regions, within which the anomalous Hall conductivity vanishes,
may occur in the supermoir\'e pattern.  (If the number $N$ of tBG moir\'e cells in a supercell of the commensurate
system is a multiple of 4, for example in the $90^\circ$ commensurate case, it is not impossible that the Fermi level
could lie within one of the subband gaps of the original moir\'e bands.)
According to the results shown in Fig. \ref{fig_phasemaps} (d)-(i),
unpolarized states are more likely in semimetal phases of nearly $60^\circ$ commensurate systems
and in $C=1$ regions in nearly $120^\circ$ commensurate systems.

\subsection{Supermoir\'e quantum anomalous Hall effect twist angle windows}
\label{sec_twistanglewindows}

Our percolation-like \cite{Trugman1983,Chalker_1988} picture of the supermoir\'e anomalous Hall
effect allows the spatial maps in Fig.~\ref{fig_phasemaps} to be interpreted using a
Landauer-b\"{u}ttiker transport picture. \cite{Buttiker1986,JingWang2013}
In this picture an overall quantized anomalous Hall conductance occurs
only when (i) a topologically nontrivial QAH phase percolates;
(ii) the edge states between phase boundaries are sufficiently localized that their coupling can be neglected.
The latter condition requires that the twist angle pair should be sufficiently close to a commensurate point
that the supermoir\'e period is large compared to the lateral localization of the edge states.
These considerations lead to the conclusion that there is a region of finite area in twist angle space surrounding
each commensurate point within which the QAH effect can occur.
Below we provide an estimate of the sizes of these twist angle windows.

We estimate the lateral localization width $\lambda$ of the edge states localized along
boundaries between topologically nontrivial and trivial phases by concentrating on the
two crossing levels and appealing to a Jackiw-Rebbi picture \cite{Jackiw1976,SuSchrieffer1979} of two-dimensional Dirac
fermions with a mass gap that varies smoothly with position.  This mapping yields
\begin{equation}
    \lambda=2\sqrt{\frac{v_F}{|\nabla_{\rr}\Delta|}},
    \label{eq_localization}
\end{equation}
where $\Delta$ is the local gap.  The typical Fermi velocity, $\hbar v_F \approx 100 \rm{meV\cdot nm}$,
was estimated from our model calculations by examining band dispersion at touching points like the one in
Fig. \ref{fig_C(d)} (c).  Similarly the rate of variation of the gap with $\dd$ is $|\nabla_{\dd}\Delta|\approx 300\rm{meV}\cdot\rm{nm}^{-1}$.
For a supermoir\'e lattice with a magnification factor $|\rr|/|\dd|=\gamma$, we have $|\nabla_{\rr}\Delta|=|\nabla_{\dd}\Delta|/\gamma$.
Quantization is accurate when the edge-isolation parameter
$\rho \equiv a_{\rm sm}/\lambda$, the ratio of gapped state size to edge state localization length,
is large.  From Eqs. (\ref{eq_ra}) and (\ref{eq_localization}) we find that when the twist angle is tuned toward a commensurate point
defined by Eq. (\ref{eq_cmsr_condition}) with $n=p+q$,
\begin{equation}
    \rho= \frac{a_{\rm sm}}{\lambda} = \frac{a_{\rm BN}}{2}\sqrt{\frac{\gamma|\nabla_{\dd}\Delta|}{ Nv_F}},
    \label{eq:edgeisolation}
\end{equation}
where $N=n^2$.  Since the magnification factor $\gamma$ depends smoothly on twist angle,
Eq. (\ref{eq:edgeisolation}) implies that edge isolation will be achieved over smaller ranges
of twist angle near higher order (larger $N$) commensuration points.
Here we have assumed that both $v_F$ and $|\nabla_{\dd}\Delta|$ retain their order of magnitude as $n$ becomes large.
The latter assumption is justified by Eq. (\ref{eq_Vj_d_dependence}) since
\[
    |\nabla_{\dd}\Delta|\sim|\nabla_{\dd}V_{\rm BN}|\sim\left|i\gggg_j\cdot(\nabla_{\dd}\dd_M)V_j(\dd)\right|\sim G_{\rm BN}V_{\rm BN},
\]
where $G_{\rm BN}$ is the magnitude of primitive reciprocal lattice vector of the hBN, which does not change with $n$.

\begin{figure}
    \centering
    \includegraphics[width=0.48\textwidth]{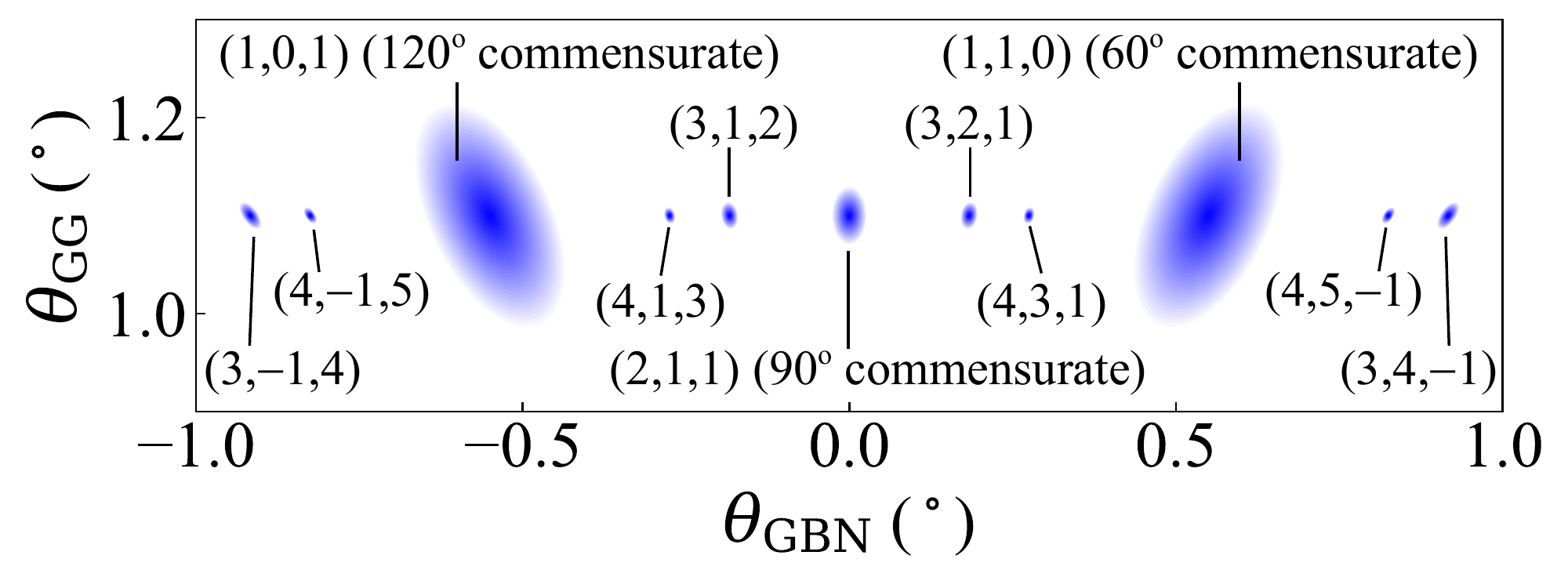}
    \caption{Twist angle windows for QAH effects according to criteria explained in the main text.
    Quantized regions are shaded blue and labeled by their $(n,p,q)$ integer triplets. The windows are larger for low-order commensurate twist angle pairs.}
    \label{fig_twistanglewindows}
\end{figure}

We adopt the practical numerical criterion that the Hall conductance is effectively quantized when the
edge isolation parameter $\rho$ exceeds 5, which according to Eq. (\ref{eq:edgeisolation}) is
equivalent to $\gamma>500n^2$ ($r_a>500n$).  From Eq. (\ref{eq_gamma}), the linear size of twist angle
window that satisfies this criterion is $\delta\theta\approx1/\gamma \approx0.1^\circ/n^2$.
The quantization windows for the series of twist angle windows up to $n=4$ are
illustrated schematically in Fig. \ref{fig_twistanglewindows}.
Within the largest two of these windows, the typical supermoir\'e period is $\sim 0.1\sim\SI{1}{\micro\metre}$,
compared to typical tBG/hBN device sizes that are up to tens of micrometers. \cite{Sharpe2019,Serlin2020,Tschirhart2020}
These considerations imply that devices can in principle be fabricated with up to tens of supermoir\'e periods on a side.

\subsection{Anomalous Hall effect of incommensurate tBG/hBN}
\label{sec_incmsr}

\begin{figure}
    \centering
    \includegraphics[width=0.48\textwidth]{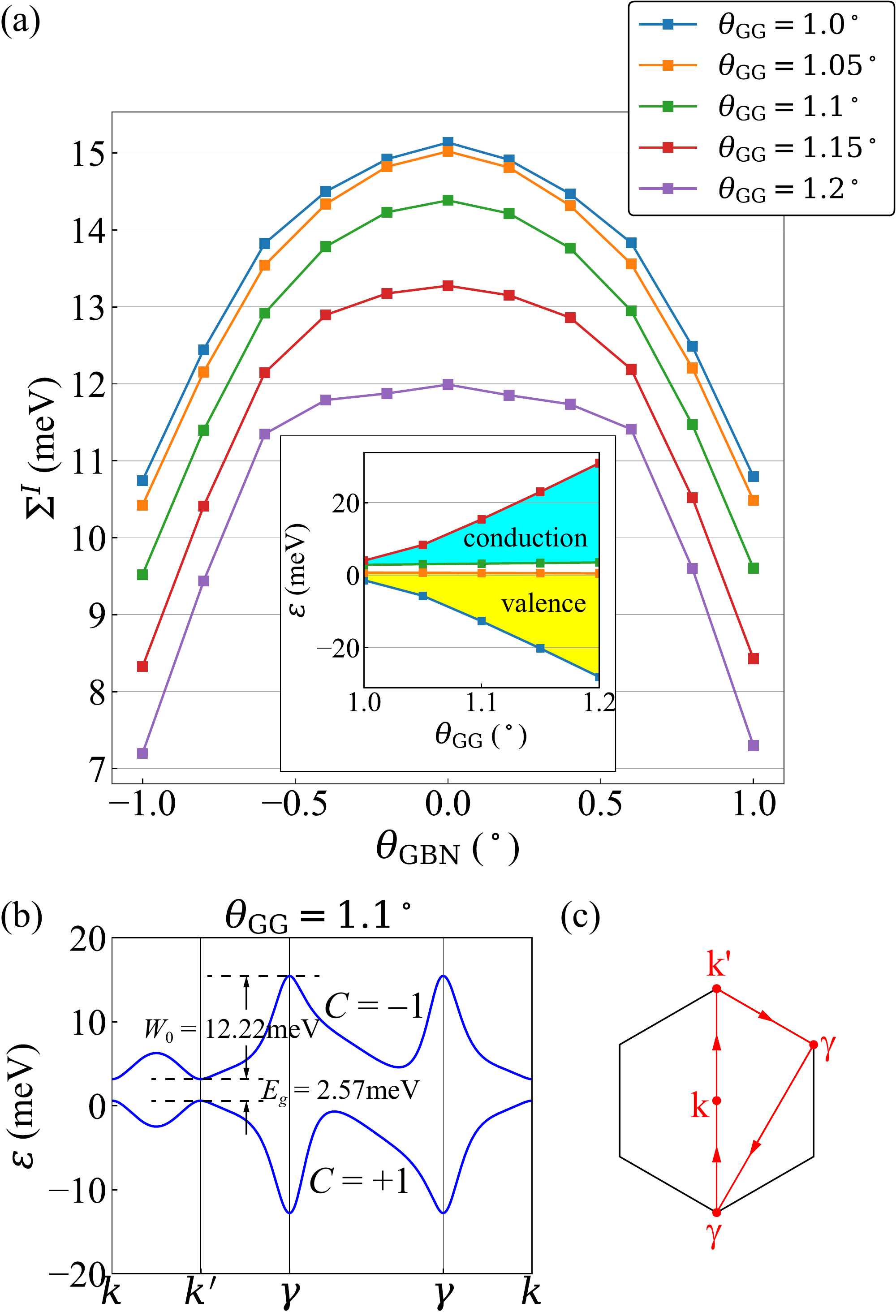}
    \caption{(a) Disorder self-energy $\Sigma^I$ due to the G1/hBN moir\'e potential {\it vs.} $\theta_{\GBN}$ for a series
    of $\theta_{\GG}$ values with $m_0=3.62\rm meV$ and interlayer potential difference $U=0$.
    Inset: the energy range of valence (yellow) and conduction (cyan) bands of near-magic angle tBG with sublattice symmetry broken by the $m_0$ term of the G1/hBN potential and the same value of $m_0$.
    (b) A sample tBG moir\'e band structure with $m_0=3.62$ meV, plotted along the red path shown in (c).
    The band Chern numbers $C$, the bandgap $E_g$, and the conduction band width $W_0$ in the absence of disorder
    are specified.}
    \label{fig_disorderwiden}
\end{figure}

In principle all twist angle pairs are close to some commensurate point,
just as all real numbers are near some rational number.
However, most of these points have extremely large $n$ and can be practically viewed as incommensurate.
In such a system, the moir\'e bands are broadened by the G1/hBN moir\'e, or split into an extremely
large number of minibands.  To roughly assess the influence of the G1/hBN moir\'e on
electronic structures in this limit we adopt a simplified picture by treating it as a
disorder potential with a scattering rate estimated using a self-consistent Born approximation:
\begin{widetext}\begin{equation}
   \tau^{-1}_{n\kk} = \frac{2 \Sigma_{n\kk}^I}{\hbar} = \frac{2\pi}{\hbar} \sum_m\sum_{j=1}^6\left|\Bra{n\kk}V_{\rm BN}\Ket{m(\kk+\gggg_j)}\right|^2\frac{1}{\pi}\frac{\Sigma_{m(\kk+\gggg_j)}^I}{(\epsilon_{m(\kk+\gggg_j)}-\epsilon_{n\kk})^2 +(\Sigma_{m(\kk+\gggg_j)}^I)^2}.
    \label{eq_Born}
\end{equation}\end{widetext}
Here $\Sigma_{n\kk}^I$ is the imaginary part of the self energy, $\Ket{n\kk}$ is the Bloch state of the $n$th band at wave vector $\kk$
and $\epsilon_{n\kk}$ is the corresponding band energy, and the $\gggg_j$'s are from the first shell of G1/hBN moir\'e pattern.
To simplify this approximation, we include only the moir\'e flat bands and
assume that the scattering rate is approximately the same for all states
by letting $\Sigma_{n\kk}^I \to \Sigma^I$ in Eq. (\ref{eq_Born}).
This yields
\begin{equation}
    \sum_{n,m=v,c}\frac{1}{N_{\kk}}\sum_{\kk\in\rm mBZ}\sum_{j=1}^6\frac{\left|\Bra{n\kk}V_{\rm BN}\Ket{m(\kk+\gggg_j)}\right|^2}{(\epsilon_{m(\kk+\gggg_j)}-\epsilon_{n\kk})^2+(\Sigma^I)^2}=1,
    \label{eq_Born_simplified}
\end{equation}
where $v$ and $c$ stand respectively for valence and conduction bands.
We solve Eq. (\ref{eq_Born_simplified}) for the disorder energy broadening $\Sigma^I$
using an $N_{\kk}=50\times50$ mesh to perform the momentum space integral and a numerical bisection method to
fix $\Sigma^I$.

Figure \ref{fig_disorderwiden} (a) shows disorder self-energy $\Sigma^I$ of the tBG bands calculated
in this way and compares them with the disorder-free band widths $W$ and gaps shown in the inset.
The disorder broadening is largest when the moir\'e bands are narrowest, as expected on the basis of density-of-states considerations, and exceeds $10 {\rm meV}$ over a broad range of twist angles.
Within a Stoner mean-field picture, spontaneous valley polarization is expected only when the moir\'e band width
is smaller than the exchange energy strength.  Assuming that the disorder self-energy $\Sigma^I$ effectively adds
to the band width, the values reported in Fig. \ref{fig_disorderwiden}
suggest that spontaneous valley polarization is unlikely in incommensurate tBG/hBN.
Since the disorder broadening effect is in any case sufficient to close the typically $2\sim3$meV band gap present
when the G1/hBN moir\'e pattern is ignored, even if present spontaneous valley polarization is unlikely
to produce a quantized anomalous Hall effect.  The property that an incommensurate tBG/hBN interaction
can be strong enough to close gaps is consistent with our findings for commensurate systems.
As also in that case a larger value for $m_0=10\rm meV$ would imply more quantum anomalous Hall
effects that are more, but still imperfectly, persistent. (see Appendix \ref{sec_app_largerm0})

\section{Discussion}
\label{sec_discuss}

When tBG/hBN devices are fabricated, $\theta_{\GG}$ can be accurately
controlled to a precision of order of $\sim 0.1^\circ$ because the two graphene sheets are
extracted from a common exfoliated single-layer crystal.\cite{Kim2016_twisttech}
This advantage is not present when aligning the graphene and hBN layers and $\theta_{\GBN}$ is therefore
far less precisely controlled.  Nominally aligned samples
may have differences in orientation in the range of $\sim \pm1^\circ$.
If the orientation angle is random within this range, the two moir\'e patterns
will generally be incommensurate and therefore, we have argued, likely to show only a
weak or zero anomalous Hall effect.
If by chance $\theta_{\GBN}$ falls into one of the twist angle windows identified
in Fig. \ref{fig_twistanglewindows}, devices are likely to exhibit a quantized Hall conductance.
Close to these twist angle windows the Hall conductance is likely to be large, but
still not quantized. This provides a possible explanation
for the fact that accurately quantized Hall conductances seem to be observed relatively rarely
in experiments on tBG/hBN.  Our expectation that the Hall conductance is more likley to be quantized
for twist angle pairs closer to a commensurate point is consistent with the experimental
observation of a quantized Hall resistance in a sample with measured twist
angles $\theta_{\GG}\approx1.15^\circ$ and $\theta_{\GBN}\approx\pm 0.6^\circ$, \cite{Serlin2020}
which is close to either the $60^\circ$ or the $120^\circ$ commensurate point depending on the sign of
$\theta_{\GBN}$, and a non-quantized Hall resistance in a sample with $\theta_{\GG}\approx 1.2^\circ$ and $\theta_{\GBN}\approx\pm0.8^\circ$, \cite{Sharpe2019} which is further from a commensurate twist-angle-pair point.

Since there will always be a difference in local lattice bonding energy per area between
regions with different values of the hBN sliding
vector $\dd$, a supermoir\'e structure will spontaneously expand
regions in which $\dd$ is close to the most energetically preferred value. \cite{Woods2014}
For samples smaller than a supermoir\'e period, this process will induce relaxation
towards a uniform phase with the energetically preferred value of $\dd$.
At present we do not know whether or not these uniform samples are more likely to be
Chern insulators, trivial insulators or semimetals.

In larger samples,
the supermoir\'e pattern can introduce intrinsic inhomogeneity at the micrometer scale.
One consequence is that the measured Hall conductance can be a device-specific quantity,
even for devices that have the same twist angles.
This scenario is consistent with the fact that in some devices the quantum anomalous
Hall effect is observed \cite{Serlin2020} for some source, drain and voltage contact choices and not for others.
The observation of domain walls \cite{Tschirhart2020} that remain pinned
even when the magnetization has apparently saturated is also consistent with device scale
inhomogeneity.  Persistent pinning might be associated with local absence of valley polarization
as discussed in Sec. \ref{sec_cmsrAH}.

The relationship we propose between commensurability and the appearance of the QAH effect in
tBG/hBN could be tested by measuring the twist-angle-pair of a nearly commensurate device
using Bragg interferometry. \cite{Nathanael2020_strain}
In this technique a high-energy electron beam with sub-moir\'e size is rastered through and diffracted by both graphene and hBN layers.
In tBG, the intensity of the Bragg disks varies with electron-injection position with moir\'e periodicity
as a result of spatially varying interference between the two graphene layers. For nearly commensurate tBG/hBN, we expect this
periodicity to be further modulated with a larger periodicity, namely the supermoir\'e, by a perturbation from the hBN layer.

The absence of an anomalous Hall effect in a large device could signal the absence of
valley polarization at any point, or a complex valley-polarization domain structure.  These circumstances
can be distinguished in principle by using nano-ARPES \cite{Aaron2012_nanoARPES,Dudin2010_nanoARPES} to separately detect
energy and momentum distribution functions in opposite valleys to see if they are different.\cite{Jihang2020_ARPES}
Valley polarization can also be measured locally by looking for valley-contrasting
optical properties.\cite{Wehling2015,Hipolito_2017,Mak2018}

\section{Summary and Conclusions}
\label{sec_summary}

Trilayer van der Waals heterojunctions have two independent relative twist angles.
We have identified a series of $(\theta_{\GG},\theta_{\GBN})$ twist-angle pairs in tBG/hBN trilayer systems
at which the graphene/graphene and graphene/hBN moir\'e periodicities are commensurate and $\theta_{\GG}$ is close
to the magic angle at which isolated tBG moir\'e bands are narrow and support strong correlation physics.
We use a non-interacting continuum model Hamiltonian that accounts for both moir\'e patterns to address the
trilayer electronic properties.  Although the active degrees of freedom are localized in the two graphene layers,
the hBN layer produces an effective external potential that includes both a position independent term, and a position-dependent
term that is often ignored.\cite{YahuiZhang2019,Nick2020,Fengcheng2020,Cecile2020,Jihang2020_Magnetic,Yahya2019,Shubhayu2020,ChengPing2020}
We find that when the position-dependent terms are retained, the band structures and Chern numbers
of commensurate trilayers change as the hBN layer is rigidly displaced by translation vector $\dd$.
When only the translationally invariant mass term are included in the Hamiltonian,
the electronic structure is $\dd$-independent, and the Chern number maps are uniform at $C=1$.
This finding proves that the role of the position-dependent terms in trilayers,
which have the periodicity of the graphene/hBN moir\'e, is essential.

Building on this result, we analyze the role of the graphene/hBN moir\'e in tBG/hBN trilayers, focusing
on their importance for the appearance or absence of the QAH effect at odd
integer moir\'e band fillings.  When the twist angle pair is close to a commensurate point,
a long-period supermoir\'e pattern is formed that can be viewed as a slow spatial variation of the
hBN translation vector $\dd$.  When analyzed using a local moir\'e band picture,
the supermoir\'e at odd integer moir\'e band filling factors is characterized
by a spatial map of distinct states, including
correlated insulating states with various Chern numbers,
semimetal states, and valley-unpolarized states.
We argue that an overall QAH state is possible only when a topologically nontrivial insulating phase
percolates and the twist angle pair is close enough to a commensurate value.
For twist angles far from commensurate points, we assume that the hBN moir\'e potential acts like a disorder potential
which we treat using a self-consistent Born approximation.  We argue that
that the anomalous Hall effect is unlikely to occur in this regime because of the disorder-induced band-broadening effect.

Our proposal can explain the experimental observation of both quantized and non-quantized anomalous Hall effects,
as well as states with no anomalous Hall effect at all, in tBG/hBN samples.
The supermoir\'e picture also provides possible interpretations of unexplained
inhomogeneities observed in some experiments that act as pinning centers of orbital ferromagnetism.
Direct verification of our proposal could be achieved by performing Bragg interferometry moir\'e structure
and transport measurements in the same sample.

Earlier experimental \cite{WangLujun2019,Finney2019,WangZihao2019,KanTing2019} and
theoretical \cite{Nicolas2019,Andelkovic2020,Amorim2018,ZiyanZhu2020_relax,ZiyanZhu2020_tTG} work has
addressed the rich electronic properties of other trilayer systems, including hBN/graphene/hBN trilayers
and twisted trilayer graphene system. This manuscript shows that the tBG/hBN trilayer system is
also an attractive platform to study bi-moir\'e electronic structures,
and to study the interplay between strong-correlations and quasiperiodicity.

\textit{Note added in proof} --- As this manuscript was being prepared we noticed a related
preprint \cite{Lin2020_instability} that identifies a series of commensurate twist angle pairs in tBG/hBN and
performed full structural relaxation calculations.  This work supports our speculation that
commensurate moir\'e structures are likely to be energetically preferred.
A second related preprint \cite{DanMao2020} has identified the two simplest examples of
mori\'e commensurability, and provides a complementary analysis of the electronic
properties of incommensurate tBG/hBN.

\begin{acknowledgements}
We acknowledge support from DOE grant DE- FG02-02ER45958 and Welch Foundation Grant F1473.
The authors acknowledge helpful interactions with David Goldhaber-Gordon, Aaron Sharpe, Andrea Young, Powel Potasz, Chunli Huang, Nemin Wei and Wei Qin. We also thank the Texas Advanced Computing Center for providing computational resources.
\end{acknowledgements}

\appendix
\renewcommand\thefigure{A\arabic{figure}}
\setcounter{figure}{0}

\section{Exact geometry of general commensurate tBG/hBN trilayers}
\label{sec_app_geometry}

\begin{figure}
    \centering
    \includegraphics[width=0.48\textwidth]{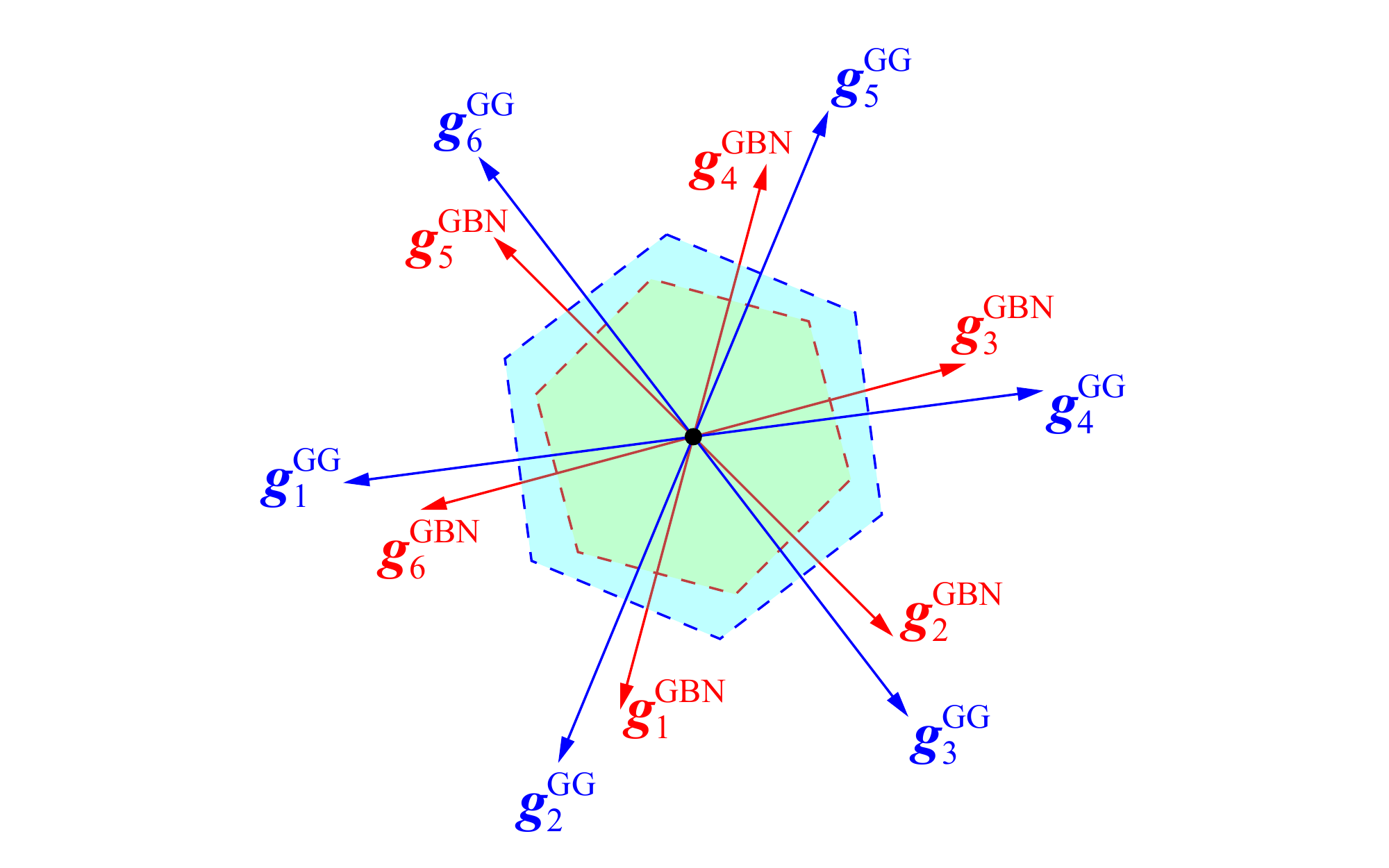}
    \caption{The primitive reciprocal lattice vectors of the G1/G2 (blue) and G1/hBN moir\'e patterns (red).}
    \label{fig_app_geometry}
\end{figure}

The commensurability of the tBG/hBN trilayer is captured by the fact that any reciprocal lattice vector of either moir\'e pattern is a linear combination of the primitive basis of a common mini- reciprocal lattice with \textit{integer} coefficients. This is equivalent to saying that any reciprocal lattice vector of one moir\'e pattern is a linear combination of the reciprocal basis of the other moir\'e pattern with \textit{rational} coefficients. According to this condition we can set
\begin{equation}
    \gggg_1^{\rm GBN}=\tilde{p}\gggg_2^{\rm GG}+\tilde{q}\gggg_3^{\rm GG}
    \label{eq_app_geometry_g}
\end{equation}
where $\gggg_j^{\rm GG}$ and $\gggg_j^{\rm GBN}$ are defined in Fig. \ref{fig_app_geometry}, $\tilde{p}$ and $\tilde{q}$ are rational numbers with the least common denominator $n$ so that $\tilde{p}=p/n$ and $\tilde{q}=q/n$. Rotating both sides of Eq. (\ref{eq_app_geometry_g}) clockwise by $90^\circ$ and scaling by $1/\sqrt{3}$ yields Eq. (\ref{eq_cmsr_condition}) in the main text.

We now solve for the exact expression of the twist angle pair $(\theta_{\GG},\theta_{\GBN})$ in terms of $(\tilde{p},\tilde{q})$. We first write Eq. (\ref{eq_cmsr_condition}) in a complex number form in which
2D vectors are represented by complex numbers whose real and imaginary parts are the two components \textit{i.e.} $\KK_1=K$, $\KK_2=Ke^{i\theta_{\GG}}$ and $\KK_{\rm BN}=Ke^{i\theta_{\GBN}}/\alpha$.
Rotation matrices are then represented by complex numbers with norm 1 \textit{i.e.} $\mR_\phi=e^{i\phi}$:
\begin{equation}
    \frac{e^{i\theta_{\GBN}}}{\alpha}-1=\left(\tilde{p}e^{i\frac{\pi}{3}}+\tilde{q}e^{i\frac{2\pi}{3}}\right)(e^{i\theta_{\GG}}-1).
    \label{eq_app_complexrelation}
\end{equation}
Adding $1$ to each side of Eq. (\ref{eq_app_complexrelation}) and then multiplying by complex conjugates
yields an equation for $\theta_{\GG}$ which has two \textit{exact} solutions modulo $2\pi$:
\begin{equation}
    \theta_{\GG}^\pm=\arccos\frac{t}{\sqrt{t^2+s^2}}\pm\arccos\frac{t+\frac{1}{2}\left(1-\frac{1}{\alpha^2}\right)}{\sqrt{t^2+s^2}},
    \label{eq_app_thetaGG}
\end{equation}
where $t=r^2+s^2-r$, $r=(\tilde{p}-\tilde{q})/2$ and $s=\sqrt{3}(\tilde{p}+\tilde{q})/2$. $\theta_{\GG}^+$ is typically not small
enough to justify the continuum models that make the use of moir\'e periodic Hamiltonians.
On the other hand $\theta_{\GG}^-$ is small since $\alpha$ is very close to 1.

By similar means we can also get an equation of $\theta_{\GBN}$ from Eq. (\ref{eq_app_complexrelation}), which has two \textit{exact} solutions modulo $2\pi$:
\begin{equation}
    \theta_{\GBN}^\pm=\arccos\frac{r-1}{\sqrt{(r-1)^2+s^2}}\pm\arccos\frac{\alpha r-\frac{1}{2}\left(\alpha+\frac{1}{\alpha}\right)}{\sqrt{(r-1)^2+s^2}}.
    \label{eq_app_thetaGBN}
\end{equation}
Again, $\theta_{\GBN}^+$ is typically not small enough to justify moir\'e band theory.

The three special cases discussed in the main text are obtained by
substituting $(\tilde{p},\tilde{q})=(1,0)$, $(0,1)$ and $(1/2,1/2)$ into Eqs. (\ref{eq_app_thetaGG}) and (\ref{eq_app_thetaGBN}),
and using $\alpha=1.017$.  We obtain:
\begin{equation}
    \theta_{\GG}^{60^\circ}=60^\circ-\arccos\left(1-\frac{1}{2\alpha^2}\right)\approx1.103^\circ, \end{equation}\begin{equation}
    \theta_{\GBN}^{60^\circ}=\frac{\theta_{\GG}^{60^\circ}}{2}=\arccos\left(\frac{1}{2\alpha}\right)-60^\circ\approx0.551^\circ, \end{equation}\begin{equation}
    \theta_{\GG}^{120^\circ}=30^\circ-\arccos\left(\frac{2}{\sqrt{3}}\left(1-\frac{1}{4\alpha^2}\right)\right)\approx1.116^\circ, \end{equation}\begin{equation}
    \theta_{\GBN}^{120^\circ}=-30^\circ+\arccos\left(\frac{1}{\sqrt{3}}\left(\alpha+\frac{1}{2\alpha}\right)\right)\approx-0.577^\circ, \end{equation}\begin{equation}
    \theta_{\GG}^{90^\circ}=\arccos\sqrt{\frac{3}{7}}-\arccos\left(\frac{1}{\sqrt{21}}\left(5-\frac{2}{\alpha^2}\right)\right)\approx1.106^\circ, \end{equation}\begin{equation}
    \theta_{\GBN}^{90^\circ}=\arccos\left(\frac{1}{\sqrt{7}}\left(\alpha+\frac{1}{\alpha}\right)\right)-\arccos\frac{2}{\sqrt{7}} \approx-0.009^\circ.
\end{equation}

\hspace*{\fill}
\section{Geometry of tBG/hBN supermoir\'e}
\label{sec_app_geometry_sm}

\begin{figure*}
    \centering
    \includegraphics[width=0.96\textwidth]{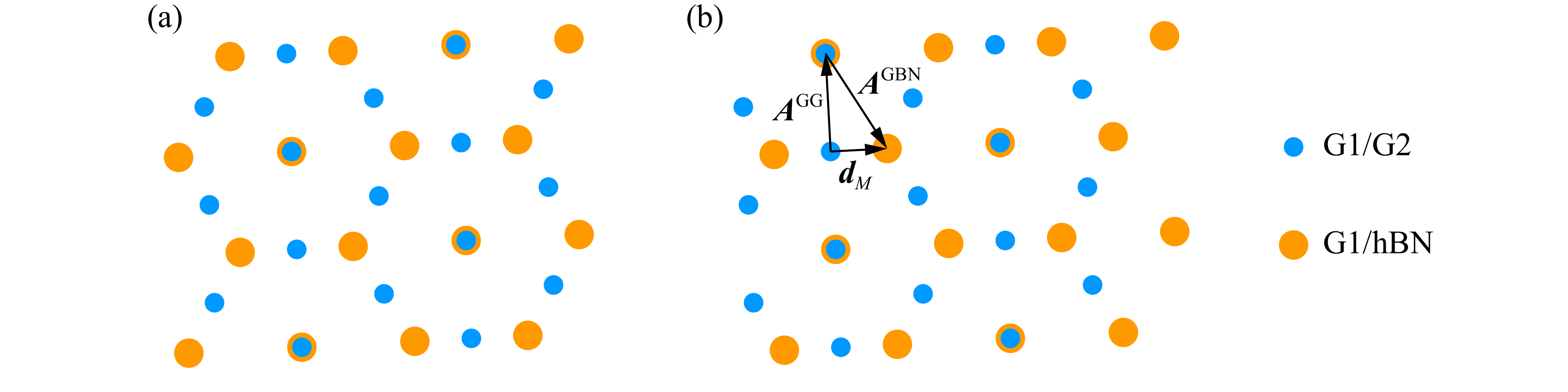}
    \caption{Local AA stacking points of the two moir\'e patterns in a $90^\circ$ commensurate system: (a) $\dd_M=0$; (b) $\dd_M$ is the half of a shortest G1/hBN moir\'e lattice vector, which is the sum of a G1/G2 moir\'e lattice vector $\AAA^{\rm GG}$ and a G1/hBN moir\'e lattice vector $\AAA^{\rm GBN}$. The two systems are identical up to a translation. The small blue and large orange dots represent the local AA stacking points of G1/G2 and G1/hBN moir\'e patterns respectively.}
    \label{fig_app_moirelattices}
\end{figure*}

For a tBG/hBN trilayer with twist angle pair $(\theta_{\GG},\theta_{\GBN})$, the two moir\'e Bravais lattices are defined by
\begin{equation}
    \begin{cases}
        \AAA^{\GG}=\left(1-\mR_{-\theta_{\GG}}\right)^{-1}\aaa \\
        \AAA^{\GBN}=\left(1-\frac{\mR_{-\theta_{\GBN}}}{\alpha}\right)^{-1}\aaa
    \end{cases}
    \label{eq_app_geometry_A}
\end{equation}
where $\aaa$ is a lattice vector of the G1 graphene layer.

We start from an $(n,p,q)$ commensurate structure with $\dd=0$, so that the two moir\'e patterns share AA stacking points at the origin,
and look for other common AA stacking points $\rr$ that satisfy
\begin{equation}
    \rr=\left(1-\mR_{-\theta_{\GG}^{npq}}\right)^{-1}\aaa_1=\left(1-\frac{\mR_{-\theta_{\GBN}^{npq}}}{\alpha}\right)^{-1}\aaa_2,
    \label{eq_app_sm_commonAA}
\end{equation}
where both $\aaa_1$ and $\aaa_2$ are G1 lattice vectors. Now we tune the twist angle pair slightly away by $(\delta\theta_{\GG},\delta\theta_{\GBN})$, and then the AA stacking points in both moir\'e patterns are shifted and their relative displacement is
\begin{equation}
    \dd_M(\rr)=\left(1-\frac{\mR_{-\theta_{\GBN}}}{\alpha}\right)^{-1}\aaa_2-\left(1-\mR_{-\theta_{\GG}}\right)^{-1}\aaa_1,
    \label{eq_app_sm_dM}
\end{equation}
where $\theta_{\GG}=\theta_{\GG}^{npq}+\delta\theta_{\GG}$ and $\theta_{\GBN}=\theta_{\GBN}^{npq}+\delta\theta_{\GBN}$.
Writing $\aaa_1$ and $\aaa_2$ in Eq. (\ref{eq_app_sm_dM}) in terms of $\rr$ using Eq. (\ref{eq_app_sm_commonAA})
yields an explicit expression for $\dd_M(\rr)$, and then an explicit expression of $\dd(\rr)$ by using Eq. (\ref{eq_dM_d}).
For small $\delta\theta_{\GG}$ and small $\delta\theta_{\GBN}$,
\begin{widetext}\begin{equation}
    \dd(\rr)=\left(\delta\theta_{\GBN}\mR_{90^\circ}-\frac{\alpha}{n}\delta\theta_{\GG}\left(p\mR_{30^\circ}+q\mR_{-30^\circ}\right) \mR_{\theta_{\GBN}^{npq}-\theta_{\GG}^{npq}}\right)\rr .
\end{equation}\end{widetext}
To obtain this expression one needs to make use of the relation
\begin{equation}
    n\left(1-\frac{\mR_{\theta_{\GBN}^{npq}}}{\alpha}\right)=\left(p\mR_{60^\circ}+q\mR_{120^\circ}\right)\left(1-\mR_{\theta_{\GG}^{npq}}\right).
    \label{eq_app_factorrelations}
\end{equation}
which can be extracted directly from Eq. (\ref{eq_cmsr_condition}).

Further approximation neglecting the difference between $\mR_{\theta_{\GG}^{npq}}$, $\mR_{\theta_{\GBN}^{npq}}$, $\alpha$ and 1 yields
\begin{equation}
    \dd(\rr)\approx\left(\delta\theta_{\GBN}\mR_{90^\circ}-\frac{1}{n}\delta\theta_{\GG}(p\mR_{30^\circ}+q\mR_{-30^\circ})\right)\rr.
    \label{eq_app_ddvszz}
\end{equation}
Take the norm of both sides of Eq. (\ref{eq_app_ddvszz}) and we get Eq. (\ref{eq_gamma}) in the main text.

To understand the factor $1/\sqrt{N}$ in Eq. (\ref{eq_ra}), we must return to the commensurate system and show that the system is invariant not only under a change of $\dd$ by a lattice vector of the hBN, but also under a change of $\dd$ by a lattice vector of a lattice that is $N$ times as dense as the hBN. We look at the shift of the position of G1/hBN moir\'e pattern, $\dd_M$, due to the change in $\dd$. The system is obviously invariant under a shift of $\dd_M$ by any G1/hBN moir\'e lattice vector $\AAA^{\GBN}$, and in fact also invariant under a shift of the G1/hBN moir\'e pattern by any G1/G2 moir\'e lattice vector $\AAA^{\GG}$, which can be understood by noticing its equivalence to a shift of the G1/G2 moir\'e pattern by $-\AAA^{\GG}$. An example is shown in Fig. \ref{fig_app_moirelattices}.
We also notice that combining Eqs. (\ref{eq_app_geometry_A}) and (\ref{eq_app_factorrelations}) (note that we are dealing with commensurate systems so $\theta_{\GG}=\theta_{\GG}^{npq}$, $\theta_{\GBN}=\theta_{\GBN}^{npq}$) yields the relation between the two moir\'e Bravais lattices:
\begin{equation}
    n\AAA^{\GG}=\left(p\mR_{-60^\circ}+q\mR_{-120^\circ}\right)\AAA^{\GBN}
\end{equation}
which is identical to the relation between the two moir\'e reciprocal lattices characterized by Eq. (\ref{eq_app_geometry_g}), up to a mirror reflection. Since this relation folds the mBZ of the G1/G2 moir\'e pattern into $1/N$ of its area, it also folds the spatial primitive cell of the G1/hBN moir\'e pattern into $1/N$ of its area. Hence we conclude that the system is invariant under a shift of $\dd_M$ by a lattice vector of a triangular lattice that is $N$ times as dense as the Bravais lattice of the G1/hBN pattern, which is equivalent to a shift of $\dd$ by a lattice vector of a lattice that is $N$ times as dense as the hBN.

\hspace*{\fill}
\section{Details of model Hamiltonian}
\label{sec_app_Hamiltonian}

The continuum model Hamiltonian of tBG \cite{Rafi2011} in one microscopic valley with a tunable interlayer potential difference $U$ is
\begin{widetext}\begin{equation}
    H_{\rm tBG}=\sum_{\kk}\left(\psi_{1\kk}^\dag\left(-\frac{U}{2}+\hbar v\ssigma_1\cdot\kk\right)\psi_{1\kk} +\psi_{2\kk}^\dag\left(\frac{U}{2}+\hbar v\ssigma_2\cdot\kk\right)\psi_{2\kk}\right) 
    +\left(\sum_{\kk}\sum_{j=1}^3\psi_{1\kk}^\dag T_j\psi_{2(\kk+\qq_j)}+\rm H.c.\right),
    \label{eq_app_HtBG}
\end{equation}\end{widetext}
where $v\ssigma_l\cdot\kk$ ($l=1,2$) is the graphene Dirac Hamiltonian of the $l$th layer, with $\ssigma_1=(\sigma^x,\sigma^y)$, $\ssigma_2=(\cos\theta_{\GG}\sigma^x-\sin\theta_{\GG}\sigma^y,\sin\theta_{\GG}\sigma^x+\cos\theta_{\GG}\sigma^y)$ and $v=10^6\rm m/s$.
\begin{equation}
    T_j=\begin{pmatrix}
        w_{AA} & e^{-i\frac{2\pi}{3}(j-1)}w_{AB} \\
        e^{i\frac{2\pi}{3}(j-1)}w_{AB} & w_{AA}
    \end{pmatrix}
\end{equation}
are the three interlayer tunneling matrices where $w_{AB}=113\rm meV$ \cite{Jeil2014} and $w_{AA}=0.8w_{AB}$. \cite{Stephen2019} The vectors $\qq_j$ are shown in Fig. \ref{fig_geometry} (b). Note that we have written the $T_j$ matrices in a convention taking a local AA-stacking point as the origin, which is different from Ref. \onlinecite{Rafi2011} where AB-stacking is taken as the origin.

The hBN layer adds to the Hamiltonian the term $V_{\rm BN}$ specified in Eq. (\ref{eq_VBN}) in main text, where the 6 transfer momenta $\gggg_j$ are defined in Fig. \ref{fig_geometry} (b) for arbitrary $\theta_{\GBN}$. The transfer matrices $V_j$ depends on $\dd$ via Eq. (\ref{eq_Vj_d_dependence}). $C_3$ symmetry requires that $V_j(0)$ has the following forms:
\begin{gather}
    V_1(0)=V_4^\dag(0)=\begin{pmatrix}
        C_0+C_z & C_{AB} \\ C_{AB} & C_0-C_z
    \end{pmatrix} \\
    V_3(0)=V_6^\dag(0)=\begin{pmatrix}
        C_0+C_z & e^{-i\frac{2\pi}{3}}C_{AB} \\ e^{i\frac{2\pi}{3}}C_{AB} & C_0-C_z
    \end{pmatrix} \\
    V_5(0)=V_2^\dag(0)=\begin{pmatrix}
        C_0+C_z & e^{i\frac{2\pi}{3}}C_{AB} \\ e^{-i\frac{2\pi}{3}}C_{AB} & C_0-C_z
    \end{pmatrix}
\end{gather}
where $C_0$, $C_z$ and $C_{AB}$ are complex values with dimension of energy. Different \textit{ab initio} results of these quantities as well as the mass term $m_0$ under various assumptions are presented in Refs. \onlinecite{Jeil2014,Jeil2015,Jeil2017}. Here we use the most realistic one, ``relaxed $\beta$'' in Ref. \onlinecite{Jeil2017}:
\begin{equation}\begin{array}{c}
    m_0=3.62\rm meV \\
    C_0=7.03e^{i(134.54^\circ)}\rm meV \\
    C_z=6.85e^{i(60.14^\circ)}\rm meV \\
    C_{AB}=12.94e^{i(-13.81^\circ)}\rm meV
\end{array}\end{equation}

\section{Results for larger mass term}
\label{sec_app_largerm0}

\begin{table*}
    \caption{Summary of percolating supermoir\'e phases of different commensurate structures under various interlayer potential difference $U$, with $m_0=10\rm meV$. S labels percolating semimetal states; X labels states with no percolating phase.}
    \begin{tabular}{|c|c|c|c|c|c|c|c|c|c|c|c|}
        \hline $U$ (meV) & $-100$ & $-80$ & $-60$ & $-40$ & $-20$ & 0 & 20 & 40 & 60 & 80 & 100 \\ \hline
        $60^\circ$ commensurate & $C=1$ & $C=1$ & $C=1$ & $C=1$ & $C=1$ & $C=1$ & $C=1$ & $C=1$ & $C=1$ & S & S \\ \hline
        $90^\circ$ commensurate & $C=1$ & $C=1$ & $C=1$ & $C=1$ & $C=1$ & $C=1$ & $C=1$ & $C=1$ & $C=1$ & X & S \\ \hline
        $120^\circ$ commensurate & $C=1$ & $C=1$ & $C=1$ & $C=1$ & $C=1$ & $C=1$ & $C=1$ & $C=1$ & $C=1$ & $C=1$ & $C=1$ \\ \hline
    \end{tabular}
    \label{table_app_percolation}
\end{table*}

\begin{figure}
    \centering
    \includegraphics[width=0.48\textwidth]{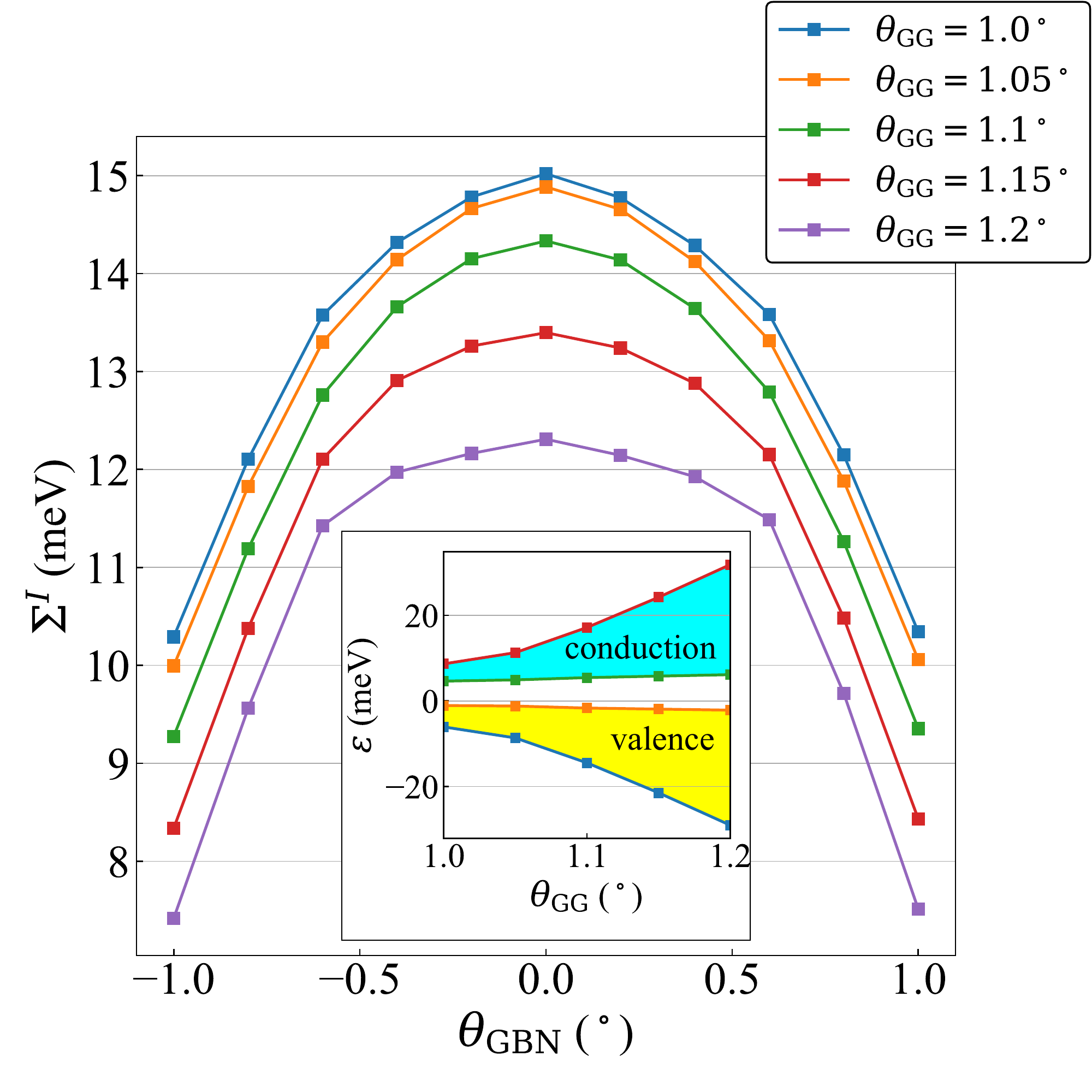}
    \caption{Disorder self-energy $\Sigma^I$ due to the G1/hBN moir\'e potential {\it vs.} $\theta_{\GBN}$ for a series
    of $\theta_{\GG}$ values with $m_0=10\rm meV$ and interlayer potential difference $U=0$.
    Inset: the energy range of valence (yellow) and conduction (cyan) bands of near-magic angle tBG with sublattice symmetry broken by the $m_0$ term of the G1/hBN potential and the same value of $m_0$.}
    \label{fig_app_disorderwiden}
\end{figure}

Table \ref{table_app_percolation} shows the percolating phase of tBG/hBN supermoir\'e structures with $m_0=10\rm meV$ and various interlayer potential difference $U$. The $C=1$ region nearly always percolates, except for very large $U$. Figure \ref{fig_app_disorderwiden} shows the estimated broadening effect $\Sigma^I$ of the periodical part of the G1/hBN moir\'e potential on the tBG bands gapped by the spatially uniform sublattice asymmetric term with $m_0=10\rm meV$. The gap increases with $\theta_{\GG}$, ranging from $\sim6\rm meV$ to $\sim8\rm meV$ in the near-magic angle regime. The broadening effect is large enough to close the gap except for relatively large $\theta_{\GG}$ and relatively large $\theta_{\GBN}$. For larger $\theta_{\GG}$ the original bandwidth $W_0$ is large, thus the full bandwidth $W\sim W_0+\Sigma^I$ is very likely to destroy the valley polarization, resulting in zero anomalous Hall conductance.
For smaller $\theta_{\GG}$ non-quantized anomalous Hall conductance is possible.

\bibliographystyle{apsrev4-1}
\bibliography{reference}

\begin{thebibliography}{72}%
\makeatletter
\providecommand \@ifxundefined [1]{%
 \@ifx{#1\undefined}
}%
\providecommand \@ifnum [1]{%
 \ifnum #1\expandafter \@firstoftwo
 \else \expandafter \@secondoftwo
 \fi
}%
\providecommand \@ifx [1]{%
 \ifx #1\expandafter \@firstoftwo
 \else \expandafter \@secondoftwo
 \fi
}%
\providecommand \natexlab [1]{#1}%
\providecommand \enquote  [1]{``#1''}%
\providecommand \bibnamefont  [1]{#1}%
\providecommand \bibfnamefont [1]{#1}%
\providecommand \citenamefont [1]{#1}%
\providecommand \href@noop [0]{\@secondoftwo}%
\providecommand \href [0]{\begingroup \@sanitize@url \@href}%
\providecommand \@href[1]{\@@startlink{#1}\@@href}%
\providecommand \@@href[1]{\endgroup#1\@@endlink}%
\providecommand \@sanitize@url [0]{\catcode `\\12\catcode `\$12\catcode
  `\&12\catcode `\#12\catcode `\^12\catcode `\_12\catcode `\%12\relax}%
\providecommand \@@startlink[1]{}%
\providecommand \@@endlink[0]{}%
\providecommand \url  [0]{\begingroup\@sanitize@url \@url }%
\providecommand \@url [1]{\endgroup\@href {#1}{\urlprefix }}%
\providecommand \urlprefix  [0]{URL }%
\providecommand \Eprint [0]{\href }%
\providecommand \doibase [0]{http://dx.doi.org/}%
\providecommand \selectlanguage [0]{\@gobble}%
\providecommand \bibinfo  [0]{\@secondoftwo}%
\providecommand \bibfield  [0]{\@secondoftwo}%
\providecommand \translation [1]{[#1]}%
\providecommand \BibitemOpen [0]{}%
\providecommand \bibitemStop [0]{}%
\providecommand \bibitemNoStop [0]{.\EOS\space}%
\providecommand \EOS [0]{\spacefactor3000\relax}%
\providecommand \BibitemShut  [1]{\csname bibitem#1\endcsname}%
\let\auto@bib@innerbib\@empty
\bibitem [{\citenamefont {Bistritzer}\ and\ \citenamefont
  {MacDonald}(2011)}]{Rafi2011}%
  \BibitemOpen
  \bibfield  {author} {\bibinfo {author} {\bibfnamefont {R.}~\bibnamefont
  {Bistritzer}}\ and\ \bibinfo {author} {\bibfnamefont {A.~H.}\ \bibnamefont
  {MacDonald}},\ }\href {\doibase 10.1073/pnas.1108174108} {\bibfield
  {journal} {\bibinfo  {journal} {Proc. Natl. Acad. Sci.}\ }\textbf {\bibinfo
  {volume} {108}},\ \bibinfo {pages} {12233} (\bibinfo {year}
  {2011})}\BibitemShut {NoStop}%
\bibitem [{\citenamefont {Cao}\ \emph {et~al.}(2018{\natexlab{a}})\citenamefont
  {Cao}, \citenamefont {Fatemi}, \citenamefont {Demir}, \citenamefont {Fang},
  \citenamefont {Tomarken}, \citenamefont {Luo}, \citenamefont
  {Sanchez-Yamagishi}, \citenamefont {Watanabe}, \citenamefont {Taniguchi},
  \citenamefont {Kaxiras}, \citenamefont {Ashoori},\ and\ \citenamefont
  {Jarillo-Herrero}}]{Cao2018Correlated}%
  \BibitemOpen
  \bibfield  {author} {\bibinfo {author} {\bibfnamefont {Y.}~\bibnamefont
  {Cao}}, \bibinfo {author} {\bibfnamefont {V.}~\bibnamefont {Fatemi}},
  \bibinfo {author} {\bibfnamefont {A.}~\bibnamefont {Demir}}, \bibinfo
  {author} {\bibfnamefont {S.}~\bibnamefont {Fang}}, \bibinfo {author}
  {\bibfnamefont {S.~L.}\ \bibnamefont {Tomarken}}, \bibinfo {author}
  {\bibfnamefont {J.~Y.}\ \bibnamefont {Luo}}, \bibinfo {author} {\bibfnamefont
  {J.~D.}\ \bibnamefont {Sanchez-Yamagishi}}, \bibinfo {author} {\bibfnamefont
  {K.}~\bibnamefont {Watanabe}}, \bibinfo {author} {\bibfnamefont
  {T.}~\bibnamefont {Taniguchi}}, \bibinfo {author} {\bibfnamefont
  {E.}~\bibnamefont {Kaxiras}}, \bibinfo {author} {\bibfnamefont {R.~C.}\
  \bibnamefont {Ashoori}}, \ and\ \bibinfo {author} {\bibfnamefont
  {P.}~\bibnamefont {Jarillo-Herrero}},\ }\href@noop {} {\bibfield  {journal}
  {\bibinfo  {journal} {Nature}\ }\textbf {\bibinfo {volume} {556}},\ \bibinfo
  {pages} {80} (\bibinfo {year} {2018}{\natexlab{a}})}\BibitemShut {NoStop}%
\bibitem [{\citenamefont {Sharpe}\ \emph {et~al.}(2019)\citenamefont {Sharpe},
  \citenamefont {Fox}, \citenamefont {Barnard}, \citenamefont {Finney},
  \citenamefont {Watanabe}, \citenamefont {Taniguchi}, \citenamefont
  {Kastner},\ and\ \citenamefont {Goldhaber-Gordon}}]{Sharpe2019}%
  \BibitemOpen
  \bibfield  {author} {\bibinfo {author} {\bibfnamefont {A.~L.}\ \bibnamefont
  {Sharpe}}, \bibinfo {author} {\bibfnamefont {E.~J.}\ \bibnamefont {Fox}},
  \bibinfo {author} {\bibfnamefont {A.~W.}\ \bibnamefont {Barnard}}, \bibinfo
  {author} {\bibfnamefont {J.}~\bibnamefont {Finney}}, \bibinfo {author}
  {\bibfnamefont {K.}~\bibnamefont {Watanabe}}, \bibinfo {author}
  {\bibfnamefont {T.}~\bibnamefont {Taniguchi}}, \bibinfo {author}
  {\bibfnamefont {M.~A.}\ \bibnamefont {Kastner}}, \ and\ \bibinfo {author}
  {\bibfnamefont {D.}~\bibnamefont {Goldhaber-Gordon}},\ }\href {\doibase
  10.1126/science.aaw3780} {\bibfield  {journal} {\bibinfo  {journal}
  {Science}\ }\textbf {\bibinfo {volume} {365}},\ \bibinfo {pages} {605}
  (\bibinfo {year} {2019})}\BibitemShut {NoStop}%
\bibitem [{\citenamefont {Serlin}\ \emph {et~al.}(2020)\citenamefont {Serlin},
  \citenamefont {Tschirhart}, \citenamefont {Polshyn}, \citenamefont {Zhang},
  \citenamefont {Zhu}, \citenamefont {Watanabe}, \citenamefont {Taniguchi},
  \citenamefont {Balents},\ and\ \citenamefont {Young}}]{Serlin2020}%
  \BibitemOpen
  \bibfield  {author} {\bibinfo {author} {\bibfnamefont {M.}~\bibnamefont
  {Serlin}}, \bibinfo {author} {\bibfnamefont {C.~L.}\ \bibnamefont
  {Tschirhart}}, \bibinfo {author} {\bibfnamefont {H.}~\bibnamefont {Polshyn}},
  \bibinfo {author} {\bibfnamefont {Y.}~\bibnamefont {Zhang}}, \bibinfo
  {author} {\bibfnamefont {J.}~\bibnamefont {Zhu}}, \bibinfo {author}
  {\bibfnamefont {K.}~\bibnamefont {Watanabe}}, \bibinfo {author}
  {\bibfnamefont {T.}~\bibnamefont {Taniguchi}}, \bibinfo {author}
  {\bibfnamefont {L.}~\bibnamefont {Balents}}, \ and\ \bibinfo {author}
  {\bibfnamefont {A.~F.}\ \bibnamefont {Young}},\ }\href {\doibase
  10.1126/science.aay5533} {\bibfield  {journal} {\bibinfo  {journal}
  {Science}\ }\textbf {\bibinfo {volume} {367}},\ \bibinfo {pages} {900}
  (\bibinfo {year} {2020})}\BibitemShut {NoStop}%
\bibitem [{\citenamefont {Nuckolls}\ \emph {et~al.}(2020)\citenamefont
  {Nuckolls}, \citenamefont {Oh}, \citenamefont {Wong}, \citenamefont {Lian},
  \citenamefont {Watanabe}, \citenamefont {Taniguchi}, \citenamefont
  {Bernevig},\ and\ \citenamefont {Yazdani}}]{Kevin2020}%
  \BibitemOpen
  \bibfield  {author} {\bibinfo {author} {\bibfnamefont {K.~P.}\ \bibnamefont
  {Nuckolls}}, \bibinfo {author} {\bibfnamefont {M.}~\bibnamefont {Oh}},
  \bibinfo {author} {\bibfnamefont {D.}~\bibnamefont {Wong}}, \bibinfo {author}
  {\bibfnamefont {B.}~\bibnamefont {Lian}}, \bibinfo {author} {\bibfnamefont
  {K.}~\bibnamefont {Watanabe}}, \bibinfo {author} {\bibfnamefont
  {T.}~\bibnamefont {Taniguchi}}, \bibinfo {author} {\bibfnamefont {B.~A.}\
  \bibnamefont {Bernevig}}, \ and\ \bibinfo {author} {\bibfnamefont
  {A.}~\bibnamefont {Yazdani}},\ }\href@noop {} {\enquote {\bibinfo {title}
  {Strongly correlated chern insulators in magic-angle twisted bilayer
  graphene},}\ } (\bibinfo {year} {2020}),\ \Eprint
  {http://arxiv.org/abs/arXiv:2007.03810} {arXiv:2007.03810} \BibitemShut
  {NoStop}%
\bibitem [{\citenamefont {Saito}\ \emph
  {et~al.}(2020{\natexlab{a}})\citenamefont {Saito}, \citenamefont {Ge},
  \citenamefont {Rademaker}, \citenamefont {Watanabe}, \citenamefont
  {Taniguchi}, \citenamefont {Abanin},\ and\ \citenamefont
  {Young}}]{Saito2020}%
  \BibitemOpen
  \bibfield  {author} {\bibinfo {author} {\bibfnamefont {Y.}~\bibnamefont
  {Saito}}, \bibinfo {author} {\bibfnamefont {J.}~\bibnamefont {Ge}}, \bibinfo
  {author} {\bibfnamefont {L.}~\bibnamefont {Rademaker}}, \bibinfo {author}
  {\bibfnamefont {K.}~\bibnamefont {Watanabe}}, \bibinfo {author}
  {\bibfnamefont {T.}~\bibnamefont {Taniguchi}}, \bibinfo {author}
  {\bibfnamefont {D.~A.}\ \bibnamefont {Abanin}}, \ and\ \bibinfo {author}
  {\bibfnamefont {A.~F.}\ \bibnamefont {Young}},\ }\href@noop {} {\enquote
  {\bibinfo {title} {Hofstadter subband ferromagnetism and symmetry broken
  chern insulators in twisted bilayer graphene},}\ } (\bibinfo {year}
  {2020}{\natexlab{a}}),\ \Eprint {http://arxiv.org/abs/arXiv:2007.06115}
  {arXiv:2007.06115} \BibitemShut {NoStop}%
\bibitem [{\citenamefont {Cao}\ \emph {et~al.}(2018{\natexlab{b}})\citenamefont
  {Cao}, \citenamefont {Fatemi}, \citenamefont {Fang}, \citenamefont
  {Watanabe}, \citenamefont {Taniguchi}, \citenamefont {Kaxiras},\ and\
  \citenamefont {Jarillo-Herrero}}]{Cao2018_SC}%
  \BibitemOpen
  \bibfield  {author} {\bibinfo {author} {\bibfnamefont {Y.}~\bibnamefont
  {Cao}}, \bibinfo {author} {\bibfnamefont {V.}~\bibnamefont {Fatemi}},
  \bibinfo {author} {\bibfnamefont {S.}~\bibnamefont {Fang}}, \bibinfo {author}
  {\bibfnamefont {K.}~\bibnamefont {Watanabe}}, \bibinfo {author}
  {\bibfnamefont {T.}~\bibnamefont {Taniguchi}}, \bibinfo {author}
  {\bibfnamefont {E.}~\bibnamefont {Kaxiras}}, \ and\ \bibinfo {author}
  {\bibfnamefont {P.}~\bibnamefont {Jarillo-Herrero}},\ }\href@noop {}
  {\bibfield  {journal} {\bibinfo  {journal} {Nature}\ }\textbf {\bibinfo
  {volume} {556}},\ \bibinfo {pages} {43} (\bibinfo {year}
  {2018}{\natexlab{b}})},\ \bibinfo {note} {article}\BibitemShut {NoStop}%
\bibitem [{\citenamefont {Lu}\ \emph {et~al.}(2019)\citenamefont {Lu},
  \citenamefont {Stepanov}, \citenamefont {Yang}, \citenamefont {Xie},
  \citenamefont {Aamir}, \citenamefont {Das}, \citenamefont {Urgell},
  \citenamefont {Watanabe}, \citenamefont {Taniguchi}, \citenamefont {Zhang},
  \citenamefont {Bachtold}, \citenamefont {MacDonald},\ and\ \citenamefont
  {Efetov}}]{Lu2019}%
  \BibitemOpen
  \bibfield  {author} {\bibinfo {author} {\bibfnamefont {X.}~\bibnamefont
  {Lu}}, \bibinfo {author} {\bibfnamefont {P.}~\bibnamefont {Stepanov}},
  \bibinfo {author} {\bibfnamefont {W.}~\bibnamefont {Yang}}, \bibinfo {author}
  {\bibfnamefont {M.}~\bibnamefont {Xie}}, \bibinfo {author} {\bibfnamefont
  {M.~A.}\ \bibnamefont {Aamir}}, \bibinfo {author} {\bibfnamefont
  {I.}~\bibnamefont {Das}}, \bibinfo {author} {\bibfnamefont {C.}~\bibnamefont
  {Urgell}}, \bibinfo {author} {\bibfnamefont {K.}~\bibnamefont {Watanabe}},
  \bibinfo {author} {\bibfnamefont {T.}~\bibnamefont {Taniguchi}}, \bibinfo
  {author} {\bibfnamefont {G.}~\bibnamefont {Zhang}}, \bibinfo {author}
  {\bibfnamefont {A.}~\bibnamefont {Bachtold}}, \bibinfo {author}
  {\bibfnamefont {A.~H.}\ \bibnamefont {MacDonald}}, \ and\ \bibinfo {author}
  {\bibfnamefont {D.~K.}\ \bibnamefont {Efetov}},\ }\href@noop {} {\bibfield
  {journal} {\bibinfo  {journal} {Nature}\ }\textbf {\bibinfo {volume} {574}},\
  \bibinfo {pages} {653} (\bibinfo {year} {2019})}\BibitemShut {NoStop}%
\bibitem [{\citenamefont {Yankowitz}\ \emph {et~al.}(2019)\citenamefont
  {Yankowitz}, \citenamefont {Chen}, \citenamefont {Polshyn}, \citenamefont
  {Zhang}, \citenamefont {Watanabe}, \citenamefont {Taniguchi}, \citenamefont
  {Graf}, \citenamefont {Young},\ and\ \citenamefont {Dean}}]{Yankowitz2019}%
  \BibitemOpen
  \bibfield  {author} {\bibinfo {author} {\bibfnamefont {M.}~\bibnamefont
  {Yankowitz}}, \bibinfo {author} {\bibfnamefont {S.}~\bibnamefont {Chen}},
  \bibinfo {author} {\bibfnamefont {H.}~\bibnamefont {Polshyn}}, \bibinfo
  {author} {\bibfnamefont {Y.}~\bibnamefont {Zhang}}, \bibinfo {author}
  {\bibfnamefont {K.}~\bibnamefont {Watanabe}}, \bibinfo {author}
  {\bibfnamefont {T.}~\bibnamefont {Taniguchi}}, \bibinfo {author}
  {\bibfnamefont {D.}~\bibnamefont {Graf}}, \bibinfo {author} {\bibfnamefont
  {A.~F.}\ \bibnamefont {Young}}, \ and\ \bibinfo {author} {\bibfnamefont
  {C.~R.}\ \bibnamefont {Dean}},\ }\href {\doibase 10.1126/science.aav1910}
  {\bibfield  {journal} {\bibinfo  {journal} {Science}\ }\textbf {\bibinfo
  {volume} {363}},\ \bibinfo {pages} {1059} (\bibinfo {year}
  {2019})}\BibitemShut {NoStop}%
\bibitem [{\citenamefont {Saito}\ \emph
  {et~al.}(2020{\natexlab{b}})\citenamefont {Saito}, \citenamefont {Ge},
  \citenamefont {Watanabe}, \citenamefont {Taniguchi},\ and\ \citenamefont
  {Young}}]{Saito2020_independent}%
  \BibitemOpen
  \bibfield  {author} {\bibinfo {author} {\bibfnamefont {Y.}~\bibnamefont
  {Saito}}, \bibinfo {author} {\bibfnamefont {J.}~\bibnamefont {Ge}}, \bibinfo
  {author} {\bibfnamefont {K.}~\bibnamefont {Watanabe}}, \bibinfo {author}
  {\bibfnamefont {T.}~\bibnamefont {Taniguchi}}, \ and\ \bibinfo {author}
  {\bibfnamefont {A.~F.}\ \bibnamefont {Young}},\ }\href@noop {} {\bibfield
  {journal} {\bibinfo  {journal} {Nature Physics}\ }\textbf {\bibinfo {volume}
  {16}},\ \bibinfo {pages} {926} (\bibinfo {year}
  {2020}{\natexlab{b}})}\BibitemShut {NoStop}%
\bibitem [{\citenamefont {Chen}\ \emph {et~al.}(2019)\citenamefont {Chen},
  \citenamefont {Sharpe}, \citenamefont {Gallagher}, \citenamefont {Rosen},
  \citenamefont {Fox}, \citenamefont {Jiang}, \citenamefont {Lyu},
  \citenamefont {Li}, \citenamefont {Watanabe}, \citenamefont {Taniguchi},
  \citenamefont {Jung}, \citenamefont {Shi}, \citenamefont {Goldhaber-Gordon},
  \citenamefont {Zhang},\ and\ \citenamefont {Wang}}]{Chen2019_SC}%
  \BibitemOpen
  \bibfield  {author} {\bibinfo {author} {\bibfnamefont {G.}~\bibnamefont
  {Chen}}, \bibinfo {author} {\bibfnamefont {A.~L.}\ \bibnamefont {Sharpe}},
  \bibinfo {author} {\bibfnamefont {P.}~\bibnamefont {Gallagher}}, \bibinfo
  {author} {\bibfnamefont {I.~T.}\ \bibnamefont {Rosen}}, \bibinfo {author}
  {\bibfnamefont {E.~J.}\ \bibnamefont {Fox}}, \bibinfo {author} {\bibfnamefont
  {L.}~\bibnamefont {Jiang}}, \bibinfo {author} {\bibfnamefont
  {B.}~\bibnamefont {Lyu}}, \bibinfo {author} {\bibfnamefont {H.}~\bibnamefont
  {Li}}, \bibinfo {author} {\bibfnamefont {K.}~\bibnamefont {Watanabe}},
  \bibinfo {author} {\bibfnamefont {T.}~\bibnamefont {Taniguchi}}, \bibinfo
  {author} {\bibfnamefont {J.}~\bibnamefont {Jung}}, \bibinfo {author}
  {\bibfnamefont {Z.}~\bibnamefont {Shi}}, \bibinfo {author} {\bibfnamefont
  {D.}~\bibnamefont {Goldhaber-Gordon}}, \bibinfo {author} {\bibfnamefont
  {Y.}~\bibnamefont {Zhang}}, \ and\ \bibinfo {author} {\bibfnamefont
  {F.}~\bibnamefont {Wang}},\ }\href@noop {} {\bibfield  {journal} {\bibinfo
  {journal} {Nature}\ }\textbf {\bibinfo {volume} {572}},\ \bibinfo {pages}
  {215} (\bibinfo {year} {2019})}\BibitemShut {NoStop}%
\bibitem [{\citenamefont {Shen}\ \emph {et~al.}(2020)\citenamefont {Shen},
  \citenamefont {Chu}, \citenamefont {Wu}, \citenamefont {Li}, \citenamefont
  {Wang}, \citenamefont {Zhao}, \citenamefont {Tang}, \citenamefont {Liu},
  \citenamefont {Tian}, \citenamefont {Watanabe}, \citenamefont {Taniguchi},
  \citenamefont {Yang}, \citenamefont {Meng}, \citenamefont {Shi},
  \citenamefont {Yazyev},\ and\ \citenamefont {Zhang}}]{Shen2020}%
  \BibitemOpen
  \bibfield  {author} {\bibinfo {author} {\bibfnamefont {C.}~\bibnamefont
  {Shen}}, \bibinfo {author} {\bibfnamefont {Y.}~\bibnamefont {Chu}}, \bibinfo
  {author} {\bibfnamefont {Q.}~\bibnamefont {Wu}}, \bibinfo {author}
  {\bibfnamefont {N.}~\bibnamefont {Li}}, \bibinfo {author} {\bibfnamefont
  {S.}~\bibnamefont {Wang}}, \bibinfo {author} {\bibfnamefont {Y.}~\bibnamefont
  {Zhao}}, \bibinfo {author} {\bibfnamefont {J.}~\bibnamefont {Tang}}, \bibinfo
  {author} {\bibfnamefont {J.}~\bibnamefont {Liu}}, \bibinfo {author}
  {\bibfnamefont {J.}~\bibnamefont {Tian}}, \bibinfo {author} {\bibfnamefont
  {K.}~\bibnamefont {Watanabe}}, \bibinfo {author} {\bibfnamefont
  {T.}~\bibnamefont {Taniguchi}}, \bibinfo {author} {\bibfnamefont
  {R.}~\bibnamefont {Yang}}, \bibinfo {author} {\bibfnamefont {Z.~Y.}\
  \bibnamefont {Meng}}, \bibinfo {author} {\bibfnamefont {D.}~\bibnamefont
  {Shi}}, \bibinfo {author} {\bibfnamefont {O.~V.}\ \bibnamefont {Yazyev}}, \
  and\ \bibinfo {author} {\bibfnamefont {G.}~\bibnamefont {Zhang}},\
  }\href@noop {} {\bibfield  {journal} {\bibinfo  {journal} {Nature Physics}\
  }\textbf {\bibinfo {volume} {16}},\ \bibinfo {pages} {520} (\bibinfo {year}
  {2020})}\BibitemShut {NoStop}%
\bibitem [{\citenamefont {Cao}\ \emph {et~al.}(2020)\citenamefont {Cao},
  \citenamefont {Rodan-Legrain}, \citenamefont {Rubies-Bigorda}, \citenamefont
  {Park}, \citenamefont {Watanabe}, \citenamefont {Taniguchi},\ and\
  \citenamefont {Jarillo-Herrero}}]{Cao2020_TBBG}%
  \BibitemOpen
  \bibfield  {author} {\bibinfo {author} {\bibfnamefont {Y.}~\bibnamefont
  {Cao}}, \bibinfo {author} {\bibfnamefont {D.}~\bibnamefont {Rodan-Legrain}},
  \bibinfo {author} {\bibfnamefont {O.}~\bibnamefont {Rubies-Bigorda}},
  \bibinfo {author} {\bibfnamefont {J.~M.}\ \bibnamefont {Park}}, \bibinfo
  {author} {\bibfnamefont {K.}~\bibnamefont {Watanabe}}, \bibinfo {author}
  {\bibfnamefont {T.}~\bibnamefont {Taniguchi}}, \ and\ \bibinfo {author}
  {\bibfnamefont {P.}~\bibnamefont {Jarillo-Herrero}},\ }\href@noop {}
  {\bibfield  {journal} {\bibinfo  {journal} {Nature}\ } (\bibinfo {year}
  {2020})}\BibitemShut {NoStop}%
\bibitem [{\citenamefont {Spanton}\ \emph {et~al.}(2018)\citenamefont
  {Spanton}, \citenamefont {Zibrov}, \citenamefont {Zhou}, \citenamefont
  {Taniguchi}, \citenamefont {Watanabe}, \citenamefont {Zaletel},\ and\
  \citenamefont {Young}}]{Spanton2018}%
  \BibitemOpen
  \bibfield  {author} {\bibinfo {author} {\bibfnamefont {E.~M.}\ \bibnamefont
  {Spanton}}, \bibinfo {author} {\bibfnamefont {A.~A.}\ \bibnamefont {Zibrov}},
  \bibinfo {author} {\bibfnamefont {H.}~\bibnamefont {Zhou}}, \bibinfo {author}
  {\bibfnamefont {T.}~\bibnamefont {Taniguchi}}, \bibinfo {author}
  {\bibfnamefont {K.}~\bibnamefont {Watanabe}}, \bibinfo {author}
  {\bibfnamefont {M.~P.}\ \bibnamefont {Zaletel}}, \ and\ \bibinfo {author}
  {\bibfnamefont {A.~F.}\ \bibnamefont {Young}},\ }\href {\doibase
  10.1126/science.aan8458} {\bibfield  {journal} {\bibinfo  {journal}
  {Science}\ }\textbf {\bibinfo {volume} {360}},\ \bibinfo {pages} {62}
  (\bibinfo {year} {2018})}\BibitemShut {NoStop}%
\bibitem [{\citenamefont {Wang}\ \emph {et~al.}(2020)\citenamefont {Wang},
  \citenamefont {Shih}, \citenamefont {Ghiotto}, \citenamefont {Xian},
  \citenamefont {Rhodes}, \citenamefont {Tan}, \citenamefont {Claassen},
  \citenamefont {Kennes}, \citenamefont {Bai}, \citenamefont {Kim},
  \citenamefont {Watanabe}, \citenamefont {Taniguchi}, \citenamefont {Zhu},
  \citenamefont {Hone}, \citenamefont {Rubio}, \citenamefont {Pasupathy},\ and\
  \citenamefont {Dean}}]{LeiWang2020}%
  \BibitemOpen
  \bibfield  {author} {\bibinfo {author} {\bibfnamefont {L.}~\bibnamefont
  {Wang}}, \bibinfo {author} {\bibfnamefont {E.-M.}\ \bibnamefont {Shih}},
  \bibinfo {author} {\bibfnamefont {A.}~\bibnamefont {Ghiotto}}, \bibinfo
  {author} {\bibfnamefont {L.}~\bibnamefont {Xian}}, \bibinfo {author}
  {\bibfnamefont {D.~A.}\ \bibnamefont {Rhodes}}, \bibinfo {author}
  {\bibfnamefont {C.}~\bibnamefont {Tan}}, \bibinfo {author} {\bibfnamefont
  {M.}~\bibnamefont {Claassen}}, \bibinfo {author} {\bibfnamefont {D.~M.}\
  \bibnamefont {Kennes}}, \bibinfo {author} {\bibfnamefont {Y.}~\bibnamefont
  {Bai}}, \bibinfo {author} {\bibfnamefont {B.}~\bibnamefont {Kim}}, \bibinfo
  {author} {\bibfnamefont {K.}~\bibnamefont {Watanabe}}, \bibinfo {author}
  {\bibfnamefont {T.}~\bibnamefont {Taniguchi}}, \bibinfo {author}
  {\bibfnamefont {X.}~\bibnamefont {Zhu}}, \bibinfo {author} {\bibfnamefont
  {J.}~\bibnamefont {Hone}}, \bibinfo {author} {\bibfnamefont {A.}~\bibnamefont
  {Rubio}}, \bibinfo {author} {\bibfnamefont {A.~N.}\ \bibnamefont
  {Pasupathy}}, \ and\ \bibinfo {author} {\bibfnamefont {C.~R.}\ \bibnamefont
  {Dean}},\ }\href@noop {} {\bibfield  {journal} {\bibinfo  {journal} {Nature
  Materials}\ }\textbf {\bibinfo {volume} {19}},\ \bibinfo {pages} {861}
  (\bibinfo {year} {2020})}\BibitemShut {NoStop}%
\bibitem [{\citenamefont {Regan}\ \emph {et~al.}(2020)\citenamefont {Regan},
  \citenamefont {Wang}, \citenamefont {Jin}, \citenamefont {Bakti~Utama},
  \citenamefont {Gao}, \citenamefont {Wei}, \citenamefont {Zhao}, \citenamefont
  {Zhao}, \citenamefont {Zhang}, \citenamefont {Yumigeta}, \citenamefont
  {Blei}, \citenamefont {Carlstr{\"o}m}, \citenamefont {Watanabe},
  \citenamefont {Taniguchi}, \citenamefont {Tongay}, \citenamefont {Crommie},
  \citenamefont {Zettl},\ and\ \citenamefont {Wang}}]{Regan2020}%
  \BibitemOpen
  \bibfield  {author} {\bibinfo {author} {\bibfnamefont {E.~C.}\ \bibnamefont
  {Regan}}, \bibinfo {author} {\bibfnamefont {D.}~\bibnamefont {Wang}},
  \bibinfo {author} {\bibfnamefont {C.}~\bibnamefont {Jin}}, \bibinfo {author}
  {\bibfnamefont {M.~I.}\ \bibnamefont {Bakti~Utama}}, \bibinfo {author}
  {\bibfnamefont {B.}~\bibnamefont {Gao}}, \bibinfo {author} {\bibfnamefont
  {X.}~\bibnamefont {Wei}}, \bibinfo {author} {\bibfnamefont {S.}~\bibnamefont
  {Zhao}}, \bibinfo {author} {\bibfnamefont {W.}~\bibnamefont {Zhao}}, \bibinfo
  {author} {\bibfnamefont {Z.}~\bibnamefont {Zhang}}, \bibinfo {author}
  {\bibfnamefont {K.}~\bibnamefont {Yumigeta}}, \bibinfo {author}
  {\bibfnamefont {M.}~\bibnamefont {Blei}}, \bibinfo {author} {\bibfnamefont
  {J.~D.}\ \bibnamefont {Carlstr{\"o}m}}, \bibinfo {author} {\bibfnamefont
  {K.}~\bibnamefont {Watanabe}}, \bibinfo {author} {\bibfnamefont
  {T.}~\bibnamefont {Taniguchi}}, \bibinfo {author} {\bibfnamefont
  {S.}~\bibnamefont {Tongay}}, \bibinfo {author} {\bibfnamefont
  {M.}~\bibnamefont {Crommie}}, \bibinfo {author} {\bibfnamefont
  {A.}~\bibnamefont {Zettl}}, \ and\ \bibinfo {author} {\bibfnamefont
  {F.}~\bibnamefont {Wang}},\ }\href@noop {} {\bibfield  {journal} {\bibinfo
  {journal} {Nature}\ }\textbf {\bibinfo {volume} {579}},\ \bibinfo {pages}
  {359} (\bibinfo {year} {2020})}\BibitemShut {NoStop}%
\bibitem [{\citenamefont {Liu}\ \emph {et~al.}(2020)\citenamefont {Liu},
  \citenamefont {Hao}, \citenamefont {Khalaf}, \citenamefont {Lee},
  \citenamefont {Ronen}, \citenamefont {Yoo}, \citenamefont {Haei~Najafabadi},
  \citenamefont {Watanabe}, \citenamefont {Taniguchi}, \citenamefont
  {Vishwanath},\ and\ \citenamefont {Kim}}]{Liu2020}%
  \BibitemOpen
  \bibfield  {author} {\bibinfo {author} {\bibfnamefont {X.}~\bibnamefont
  {Liu}}, \bibinfo {author} {\bibfnamefont {Z.}~\bibnamefont {Hao}}, \bibinfo
  {author} {\bibfnamefont {E.}~\bibnamefont {Khalaf}}, \bibinfo {author}
  {\bibfnamefont {J.~Y.}\ \bibnamefont {Lee}}, \bibinfo {author} {\bibfnamefont
  {Y.}~\bibnamefont {Ronen}}, \bibinfo {author} {\bibfnamefont
  {H.}~\bibnamefont {Yoo}}, \bibinfo {author} {\bibfnamefont {D.}~\bibnamefont
  {Haei~Najafabadi}}, \bibinfo {author} {\bibfnamefont {K.}~\bibnamefont
  {Watanabe}}, \bibinfo {author} {\bibfnamefont {T.}~\bibnamefont {Taniguchi}},
  \bibinfo {author} {\bibfnamefont {A.}~\bibnamefont {Vishwanath}}, \ and\
  \bibinfo {author} {\bibfnamefont {P.}~\bibnamefont {Kim}},\ }\href@noop {}
  {\bibfield  {journal} {\bibinfo  {journal} {Nature}\ }\textbf {\bibinfo
  {volume} {583}},\ \bibinfo {pages} {221} (\bibinfo {year}
  {2020})}\BibitemShut {NoStop}%
\bibitem [{\citenamefont {Tschirhart}\ \emph {et~al.}(2020)\citenamefont
  {Tschirhart}, \citenamefont {Serlin}, \citenamefont {Polshyn}, \citenamefont
  {Shragai}, \citenamefont {Xia}, \citenamefont {Zhu}, \citenamefont {Zhang},
  \citenamefont {Watanabe}, \citenamefont {Taniguchi}, \citenamefont {Huber},\
  and\ \citenamefont {Young}}]{Tschirhart2020}%
  \BibitemOpen
  \bibfield  {author} {\bibinfo {author} {\bibfnamefont {C.~L.}\ \bibnamefont
  {Tschirhart}}, \bibinfo {author} {\bibfnamefont {M.}~\bibnamefont {Serlin}},
  \bibinfo {author} {\bibfnamefont {H.}~\bibnamefont {Polshyn}}, \bibinfo
  {author} {\bibfnamefont {A.}~\bibnamefont {Shragai}}, \bibinfo {author}
  {\bibfnamefont {Z.}~\bibnamefont {Xia}}, \bibinfo {author} {\bibfnamefont
  {J.}~\bibnamefont {Zhu}}, \bibinfo {author} {\bibfnamefont {Y.}~\bibnamefont
  {Zhang}}, \bibinfo {author} {\bibfnamefont {K.}~\bibnamefont {Watanabe}},
  \bibinfo {author} {\bibfnamefont {T.}~\bibnamefont {Taniguchi}}, \bibinfo
  {author} {\bibfnamefont {M.~E.}\ \bibnamefont {Huber}}, \ and\ \bibinfo
  {author} {\bibfnamefont {A.~F.}\ \bibnamefont {Young}},\ }\href@noop {}
  {\enquote {\bibinfo {title} {Imaging orbital ferromagnetism in a moir\'e
  chern insulator},}\ } (\bibinfo {year} {2020}),\ \Eprint
  {http://arxiv.org/abs/arXiv:2006.08053} {arXiv:2006.08053} \BibitemShut
  {NoStop}%
\bibitem [{\citenamefont {Xie}\ and\ \citenamefont
  {MacDonald}(2020)}]{Ming2018}%
  \BibitemOpen
  \bibfield  {author} {\bibinfo {author} {\bibfnamefont {M.}~\bibnamefont
  {Xie}}\ and\ \bibinfo {author} {\bibfnamefont {A.~H.}\ \bibnamefont
  {MacDonald}},\ }\href@noop {} {\bibfield  {journal} {\bibinfo  {journal}
  {Phys. Rev. Lett.}\ }\textbf {\bibinfo {volume} {124}},\ \bibinfo {pages}
  {097601} (\bibinfo {year} {2020})}\BibitemShut {NoStop}%
\bibitem [{\citenamefont {Po}\ \emph {et~al.}(2018)\citenamefont {Po},
  \citenamefont {Zou}, \citenamefont {Vishwanath},\ and\ \citenamefont
  {Senthil}}]{HoiChun2018}%
  \BibitemOpen
  \bibfield  {author} {\bibinfo {author} {\bibfnamefont {H.~C.}\ \bibnamefont
  {Po}}, \bibinfo {author} {\bibfnamefont {L.}~\bibnamefont {Zou}}, \bibinfo
  {author} {\bibfnamefont {A.}~\bibnamefont {Vishwanath}}, \ and\ \bibinfo
  {author} {\bibfnamefont {T.}~\bibnamefont {Senthil}},\ }\href@noop {}
  {\bibfield  {journal} {\bibinfo  {journal} {Phys. Rev. X}\ }\textbf {\bibinfo
  {volume} {8}},\ \bibinfo {pages} {031089} (\bibinfo {year}
  {2018})}\BibitemShut {NoStop}%
\bibitem [{\citenamefont {Zhang}\ \emph
  {et~al.}(2019{\natexlab{a}})\citenamefont {Zhang}, \citenamefont {Mao},
  \citenamefont {Cao}, \citenamefont {Jarillo-Herrero},\ and\ \citenamefont
  {Senthil}}]{Yahui2019_flatChern}%
  \BibitemOpen
  \bibfield  {author} {\bibinfo {author} {\bibfnamefont {Y.-H.}\ \bibnamefont
  {Zhang}}, \bibinfo {author} {\bibfnamefont {D.}~\bibnamefont {Mao}}, \bibinfo
  {author} {\bibfnamefont {Y.}~\bibnamefont {Cao}}, \bibinfo {author}
  {\bibfnamefont {P.}~\bibnamefont {Jarillo-Herrero}}, \ and\ \bibinfo {author}
  {\bibfnamefont {T.}~\bibnamefont {Senthil}},\ }\href {\doibase
  10.1103/PhysRevB.99.075127} {\bibfield  {journal} {\bibinfo  {journal} {Phys.
  Rev. B}\ }\textbf {\bibinfo {volume} {99}},\ \bibinfo {pages} {075127}
  (\bibinfo {year} {2019}{\natexlab{a}})}\BibitemShut {NoStop}%
\bibitem [{\citenamefont {Zhang}\ \emph
  {et~al.}(2019{\natexlab{b}})\citenamefont {Zhang}, \citenamefont {Mao},\ and\
  \citenamefont {Senthil}}]{YahuiZhang2019}%
  \BibitemOpen
  \bibfield  {author} {\bibinfo {author} {\bibfnamefont {Y.-H.}\ \bibnamefont
  {Zhang}}, \bibinfo {author} {\bibfnamefont {D.}~\bibnamefont {Mao}}, \ and\
  \bibinfo {author} {\bibfnamefont {T.}~\bibnamefont {Senthil}},\ }\href@noop
  {} {\bibfield  {journal} {\bibinfo  {journal} {Phys. Rev. Research}\ }\textbf
  {\bibinfo {volume} {1}},\ \bibinfo {pages} {033126} (\bibinfo {year}
  {2019}{\natexlab{b}})}\BibitemShut {NoStop}%
\bibitem [{\citenamefont {Bultinck}\ \emph {et~al.}(2020)\citenamefont
  {Bultinck}, \citenamefont {Chatterjee},\ and\ \citenamefont
  {Zaletel}}]{Nick2020}%
  \BibitemOpen
  \bibfield  {author} {\bibinfo {author} {\bibfnamefont {N.}~\bibnamefont
  {Bultinck}}, \bibinfo {author} {\bibfnamefont {S.}~\bibnamefont
  {Chatterjee}}, \ and\ \bibinfo {author} {\bibfnamefont {M.~P.}\ \bibnamefont
  {Zaletel}},\ }\href@noop {} {\bibfield  {journal} {\bibinfo  {journal} {Phys.
  Rev. Lett.}\ }\textbf {\bibinfo {volume} {124}},\ \bibinfo {pages} {166601}
  (\bibinfo {year} {2020})}\BibitemShut {NoStop}%
\bibitem [{\citenamefont {Moon}\ and\ \citenamefont
  {Koshino}(2014)}]{Moon2014}%
  \BibitemOpen
  \bibfield  {author} {\bibinfo {author} {\bibfnamefont {P.}~\bibnamefont
  {Moon}}\ and\ \bibinfo {author} {\bibfnamefont {M.}~\bibnamefont {Koshino}},\
  }\href@noop {} {\bibfield  {journal} {\bibinfo  {journal} {Phys. Rev. B}\
  }\textbf {\bibinfo {volume} {90}},\ \bibinfo {pages} {155406} (\bibinfo
  {year} {2014})}\BibitemShut {NoStop}%
\bibitem [{\citenamefont {Wallbank}\ \emph {et~al.}(2015)\citenamefont
  {Wallbank}, \citenamefont {Mucha-Kruczy\'nski}, \citenamefont {Chen},\ and\
  \citenamefont {Fal'ko}}]{Wallbank2015}%
  \BibitemOpen
  \bibfield  {author} {\bibinfo {author} {\bibfnamefont {J.~R.}\ \bibnamefont
  {Wallbank}}, \bibinfo {author} {\bibfnamefont {M.}~\bibnamefont
  {Mucha-Kruczy\'nski}}, \bibinfo {author} {\bibfnamefont {X.}~\bibnamefont
  {Chen}}, \ and\ \bibinfo {author} {\bibfnamefont {V.~I.}\ \bibnamefont
  {Fal'ko}},\ }\href@noop {} {\bibfield  {journal} {\bibinfo  {journal}
  {Annalen der Physik}\ }\textbf {\bibinfo {volume} {527}},\ \bibinfo {pages}
  {359} (\bibinfo {year} {2015})}\BibitemShut {NoStop}%
\bibitem [{\citenamefont {Jung}\ \emph {et~al.}(2014)\citenamefont {Jung},
  \citenamefont {Raoux}, \citenamefont {Qiao},\ and\ \citenamefont
  {MacDonald}}]{Jeil2014}%
  \BibitemOpen
  \bibfield  {author} {\bibinfo {author} {\bibfnamefont {J.}~\bibnamefont
  {Jung}}, \bibinfo {author} {\bibfnamefont {A.}~\bibnamefont {Raoux}},
  \bibinfo {author} {\bibfnamefont {Z.}~\bibnamefont {Qiao}}, \ and\ \bibinfo
  {author} {\bibfnamefont {A.~H.}\ \bibnamefont {MacDonald}},\ }\href@noop {}
  {\bibfield  {journal} {\bibinfo  {journal} {Phys. Rev. B}\ }\textbf {\bibinfo
  {volume} {89}},\ \bibinfo {pages} {205414} (\bibinfo {year}
  {2014})}\BibitemShut {NoStop}%
\bibitem [{\citenamefont {Jung}\ \emph {et~al.}(2015)\citenamefont {Jung},
  \citenamefont {DaSilva}, \citenamefont {MacDonald},\ and\ \citenamefont
  {Adam}}]{Jeil2015}%
  \BibitemOpen
  \bibfield  {author} {\bibinfo {author} {\bibfnamefont {J.}~\bibnamefont
  {Jung}}, \bibinfo {author} {\bibfnamefont {A.~M.}\ \bibnamefont {DaSilva}},
  \bibinfo {author} {\bibfnamefont {A.~H.}\ \bibnamefont {MacDonald}}, \ and\
  \bibinfo {author} {\bibfnamefont {S.}~\bibnamefont {Adam}},\ }\href@noop {}
  {\bibfield  {journal} {\bibinfo  {journal} {Nature Communications}\ }\textbf
  {\bibinfo {volume} {6}},\ \bibinfo {pages} {6308} (\bibinfo {year}
  {2015})}\BibitemShut {NoStop}%
\bibitem [{\citenamefont {Jung}\ \emph {et~al.}(2017)\citenamefont {Jung},
  \citenamefont {Laksono}, \citenamefont {DaSilva}, \citenamefont {MacDonald},
  \citenamefont {Mucha-Kruczy\ifmmode~\acute{n}\else \'{n}\fi{}ski},\ and\
  \citenamefont {Adam}}]{Jeil2017}%
  \BibitemOpen
  \bibfield  {author} {\bibinfo {author} {\bibfnamefont {J.}~\bibnamefont
  {Jung}}, \bibinfo {author} {\bibfnamefont {E.}~\bibnamefont {Laksono}},
  \bibinfo {author} {\bibfnamefont {A.~M.}\ \bibnamefont {DaSilva}}, \bibinfo
  {author} {\bibfnamefont {A.~H.}\ \bibnamefont {MacDonald}}, \bibinfo {author}
  {\bibfnamefont {M.}~\bibnamefont {Mucha-Kruczy\ifmmode~\acute{n}\else
  \'{n}\fi{}ski}}, \ and\ \bibinfo {author} {\bibfnamefont {S.}~\bibnamefont
  {Adam}},\ }\href@noop {} {\bibfield  {journal} {\bibinfo  {journal} {Phys.
  Rev. B}\ }\textbf {\bibinfo {volume} {96}},\ \bibinfo {pages} {085442}
  (\bibinfo {year} {2017})}\BibitemShut {NoStop}%
\bibitem [{\citenamefont {Lin}\ and\ \citenamefont {Ni}(2019)}]{Lin2019}%
  \BibitemOpen
  \bibfield  {author} {\bibinfo {author} {\bibfnamefont {X.}~\bibnamefont
  {Lin}}\ and\ \bibinfo {author} {\bibfnamefont {J.}~\bibnamefont {Ni}},\
  }\href@noop {} {\bibfield  {journal} {\bibinfo  {journal} {Phys. Rev. B}\
  }\textbf {\bibinfo {volume} {100}},\ \bibinfo {pages} {195413} (\bibinfo
  {year} {2019})}\BibitemShut {NoStop}%
\bibitem [{\citenamefont {Zhu}\ \emph {et~al.}(2020{\natexlab{a}})\citenamefont
  {Zhu}, \citenamefont {Carr}, \citenamefont {Massatt}, \citenamefont
  {Luskin},\ and\ \citenamefont {Kaxiras}}]{ZiyanZhu2020_tTG}%
  \BibitemOpen
  \bibfield  {author} {\bibinfo {author} {\bibfnamefont {Z.}~\bibnamefont
  {Zhu}}, \bibinfo {author} {\bibfnamefont {S.}~\bibnamefont {Carr}}, \bibinfo
  {author} {\bibfnamefont {D.}~\bibnamefont {Massatt}}, \bibinfo {author}
  {\bibfnamefont {M.}~\bibnamefont {Luskin}}, \ and\ \bibinfo {author}
  {\bibfnamefont {E.}~\bibnamefont {Kaxiras}},\ }\href@noop {} {\bibfield
  {journal} {\bibinfo  {journal} {Phys. Rev. Lett.}\ }\textbf {\bibinfo
  {volume} {125}},\ \bibinfo {pages} {116404} (\bibinfo {year}
  {2020}{\natexlab{a}})}\BibitemShut {NoStop}%
\bibitem [{\citenamefont {Wu}\ and\ \citenamefont
  {Das~Sarma}(2020)}]{Fengcheng2020}%
  \BibitemOpen
  \bibfield  {author} {\bibinfo {author} {\bibfnamefont {F.}~\bibnamefont
  {Wu}}\ and\ \bibinfo {author} {\bibfnamefont {S.}~\bibnamefont {Das~Sarma}},\
  }\href@noop {} {\bibfield  {journal} {\bibinfo  {journal} {Phys. Rev. Lett.}\
  }\textbf {\bibinfo {volume} {124}},\ \bibinfo {pages} {046403} (\bibinfo
  {year} {2020})}\BibitemShut {NoStop}%
\bibitem [{\citenamefont {Repellin}\ \emph {et~al.}(2020)\citenamefont
  {Repellin}, \citenamefont {Dong}, \citenamefont {Zhang},\ and\ \citenamefont
  {Senthil}}]{Cecile2020}%
  \BibitemOpen
  \bibfield  {author} {\bibinfo {author} {\bibfnamefont {C.}~\bibnamefont
  {Repellin}}, \bibinfo {author} {\bibfnamefont {Z.}~\bibnamefont {Dong}},
  \bibinfo {author} {\bibfnamefont {Y.-H.}\ \bibnamefont {Zhang}}, \ and\
  \bibinfo {author} {\bibfnamefont {T.}~\bibnamefont {Senthil}},\ }\href@noop
  {} {\bibfield  {journal} {\bibinfo  {journal} {Phys. Rev. Lett.}\ }\textbf
  {\bibinfo {volume} {124}},\ \bibinfo {pages} {187601} (\bibinfo {year}
  {2020})}\BibitemShut {NoStop}%
\bibitem [{\citenamefont {Zhu}\ \emph {et~al.}(2020{\natexlab{b}})\citenamefont
  {Zhu}, \citenamefont {Su},\ and\ \citenamefont
  {MacDonald}}]{Jihang2020_Magnetic}%
  \BibitemOpen
  \bibfield  {author} {\bibinfo {author} {\bibfnamefont {J.}~\bibnamefont
  {Zhu}}, \bibinfo {author} {\bibfnamefont {J.-J.}\ \bibnamefont {Su}}, \ and\
  \bibinfo {author} {\bibfnamefont {A.~H.}\ \bibnamefont {MacDonald}},\
  }\href@noop {} {\enquote {\bibinfo {title} {The curious magnetic properties
  of orbital chern insulators},}\ } (\bibinfo {year} {2020}{\natexlab{b}}),\
  \Eprint {http://arxiv.org/abs/arXiv:2001.05084} {arXiv:2001.05084}
  \BibitemShut {NoStop}%
\bibitem [{\citenamefont {Alavirad}\ and\ \citenamefont
  {Sau}(2019)}]{Yahya2019}%
  \BibitemOpen
  \bibfield  {author} {\bibinfo {author} {\bibfnamefont {Y.}~\bibnamefont
  {Alavirad}}\ and\ \bibinfo {author} {\bibfnamefont {J.~D.}\ \bibnamefont
  {Sau}},\ }\href@noop {} {\enquote {\bibinfo {title} {Ferromagnetism and its
  stability from the one-magnon spectrum in twisted bilayer graphene},}\ }
  (\bibinfo {year} {2019}),\ \Eprint {http://arxiv.org/abs/arXiv:1907.13633}
  {arXiv:1907.13633} \BibitemShut {NoStop}%
\bibitem [{\citenamefont {Chatterjee}\ \emph {et~al.}(2020)\citenamefont
  {Chatterjee}, \citenamefont {Bultinck},\ and\ \citenamefont
  {Zaletel}}]{Shubhayu2020}%
  \BibitemOpen
  \bibfield  {author} {\bibinfo {author} {\bibfnamefont {S.}~\bibnamefont
  {Chatterjee}}, \bibinfo {author} {\bibfnamefont {N.}~\bibnamefont
  {Bultinck}}, \ and\ \bibinfo {author} {\bibfnamefont {M.~P.}\ \bibnamefont
  {Zaletel}},\ }\href@noop {} {\bibfield  {journal} {\bibinfo  {journal} {Phys.
  Rev. B}\ }\textbf {\bibinfo {volume} {101}},\ \bibinfo {pages} {165141}
  (\bibinfo {year} {2020})}\BibitemShut {NoStop}%
\bibitem [{\citenamefont {Zhang}\ \emph {et~al.}(2020)\citenamefont {Zhang},
  \citenamefont {Xiao}, \citenamefont {Zhou}, \citenamefont {Hu}, \citenamefont
  {Xie}, \citenamefont {Yan},\ and\ \citenamefont {Law}}]{ChengPing2020}%
  \BibitemOpen
  \bibfield  {author} {\bibinfo {author} {\bibfnamefont {C.-P.}\ \bibnamefont
  {Zhang}}, \bibinfo {author} {\bibfnamefont {J.}~\bibnamefont {Xiao}},
  \bibinfo {author} {\bibfnamefont {B.~T.}\ \bibnamefont {Zhou}}, \bibinfo
  {author} {\bibfnamefont {J.-X.}\ \bibnamefont {Hu}}, \bibinfo {author}
  {\bibfnamefont {Y.-M.}\ \bibnamefont {Xie}}, \bibinfo {author} {\bibfnamefont
  {B.}~\bibnamefont {Yan}}, \ and\ \bibinfo {author} {\bibfnamefont {K.~T.}\
  \bibnamefont {Law}},\ }\href@noop {} {\enquote {\bibinfo {title} {Giant
  nonlinear hall effect in strained twisted bilayer graphene},}\ } (\bibinfo
  {year} {2020}),\ \Eprint {http://arxiv.org/abs/arXiv:2010.08333}
  {arXiv:2010.08333} \BibitemShut {NoStop}%
\bibitem [{\citenamefont {Cea}\ \emph {et~al.}(2020)\citenamefont {Cea},
  \citenamefont {Pantale\'on},\ and\ \citenamefont {Guinea}}]{Cea2020}%
  \BibitemOpen
  \bibfield  {author} {\bibinfo {author} {\bibfnamefont {T.}~\bibnamefont
  {Cea}}, \bibinfo {author} {\bibfnamefont {P.~A.}\ \bibnamefont
  {Pantale\'on}}, \ and\ \bibinfo {author} {\bibfnamefont {F.}~\bibnamefont
  {Guinea}},\ }\href@noop {} {\bibfield  {journal} {\bibinfo  {journal} {Phys.
  Rev. B}\ }\textbf {\bibinfo {volume} {102}},\ \bibinfo {pages} {155136}
  (\bibinfo {year} {2020})}\BibitemShut {NoStop}%
\bibitem [{\citenamefont {Lin}\ and\ \citenamefont {Ni}(2020)}]{Lin2020}%
  \BibitemOpen
  \bibfield  {author} {\bibinfo {author} {\bibfnamefont {X.}~\bibnamefont
  {Lin}}\ and\ \bibinfo {author} {\bibfnamefont {J.}~\bibnamefont {Ni}},\
  }\href@noop {} {\bibfield  {journal} {\bibinfo  {journal} {Phys. Rev. B}\
  }\textbf {\bibinfo {volume} {102}},\ \bibinfo {pages} {035441} (\bibinfo
  {year} {2020})}\BibitemShut {NoStop}%
\bibitem [{\citenamefont {Wang}\ \emph
  {et~al.}(2019{\natexlab{a}})\citenamefont {Wang}, \citenamefont {Wang},
  \citenamefont {Yin}, \citenamefont {T{\'o}v{\'a}ri}, \citenamefont {Yang},
  \citenamefont {Lin}, \citenamefont {Holwill}, \citenamefont {Birkbeck},
  \citenamefont {Perello}, \citenamefont {Xu}, \citenamefont {Zultak},
  \citenamefont {Gorbachev}, \citenamefont {Kretinin}, \citenamefont
  {Taniguchi}, \citenamefont {Watanabe}, \citenamefont {Morozov}, \citenamefont
  {An{\dj}elkovi{\'c}}, \citenamefont {Milovanovi{\'c}}, \citenamefont
  {Covaci}, \citenamefont {Peeters}, \citenamefont {Mishchenko}, \citenamefont
  {Geim}, \citenamefont {Novoselov}, \citenamefont {Fal{\textquoteright}ko},
  \citenamefont {Knothe},\ and\ \citenamefont {Woods}}]{WangZihao2019}%
  \BibitemOpen
  \bibfield  {author} {\bibinfo {author} {\bibfnamefont {Z.}~\bibnamefont
  {Wang}}, \bibinfo {author} {\bibfnamefont {Y.~B.}\ \bibnamefont {Wang}},
  \bibinfo {author} {\bibfnamefont {J.}~\bibnamefont {Yin}}, \bibinfo {author}
  {\bibfnamefont {E.}~\bibnamefont {T{\'o}v{\'a}ri}}, \bibinfo {author}
  {\bibfnamefont {Y.}~\bibnamefont {Yang}}, \bibinfo {author} {\bibfnamefont
  {L.}~\bibnamefont {Lin}}, \bibinfo {author} {\bibfnamefont {M.}~\bibnamefont
  {Holwill}}, \bibinfo {author} {\bibfnamefont {J.}~\bibnamefont {Birkbeck}},
  \bibinfo {author} {\bibfnamefont {D.~J.}\ \bibnamefont {Perello}}, \bibinfo
  {author} {\bibfnamefont {S.}~\bibnamefont {Xu}}, \bibinfo {author}
  {\bibfnamefont {J.}~\bibnamefont {Zultak}}, \bibinfo {author} {\bibfnamefont
  {R.~V.}\ \bibnamefont {Gorbachev}}, \bibinfo {author} {\bibfnamefont {A.~V.}\
  \bibnamefont {Kretinin}}, \bibinfo {author} {\bibfnamefont {T.}~\bibnamefont
  {Taniguchi}}, \bibinfo {author} {\bibfnamefont {K.}~\bibnamefont {Watanabe}},
  \bibinfo {author} {\bibfnamefont {S.~V.}\ \bibnamefont {Morozov}}, \bibinfo
  {author} {\bibfnamefont {M.}~\bibnamefont {An{\dj}elkovi{\'c}}}, \bibinfo
  {author} {\bibfnamefont {S.~P.}\ \bibnamefont {Milovanovi{\'c}}}, \bibinfo
  {author} {\bibfnamefont {L.}~\bibnamefont {Covaci}}, \bibinfo {author}
  {\bibfnamefont {F.~M.}\ \bibnamefont {Peeters}}, \bibinfo {author}
  {\bibfnamefont {A.}~\bibnamefont {Mishchenko}}, \bibinfo {author}
  {\bibfnamefont {A.~K.}\ \bibnamefont {Geim}}, \bibinfo {author}
  {\bibfnamefont {K.~S.}\ \bibnamefont {Novoselov}}, \bibinfo {author}
  {\bibfnamefont {V.~I.}\ \bibnamefont {Fal{\textquoteright}ko}}, \bibinfo
  {author} {\bibfnamefont {A.}~\bibnamefont {Knothe}}, \ and\ \bibinfo {author}
  {\bibfnamefont {C.~R.}\ \bibnamefont {Woods}},\ }\href {\doibase
  10.1126/sciadv.aay8897} {\bibfield  {journal} {\bibinfo  {journal} {Science
  Advances}\ }\textbf {\bibinfo {volume} {5}} (\bibinfo {year}
  {2019}{\natexlab{a}}),\ 10.1126/sciadv.aay8897}\BibitemShut {NoStop}%
\bibitem [{\citenamefont {Leconte}\ and\ \citenamefont
  {Jung}(2019)}]{Nicolas2019}%
  \BibitemOpen
  \bibfield  {author} {\bibinfo {author} {\bibfnamefont {N.}~\bibnamefont
  {Leconte}}\ and\ \bibinfo {author} {\bibfnamefont {J.}~\bibnamefont {Jung}},\
  }\href@noop {} {\enquote {\bibinfo {title} {Commensurate and incommensurate
  double moire interference in graphene encapsulated by hexagonal boron
  nitride},}\ } (\bibinfo {year} {2019}),\ \Eprint
  {http://arxiv.org/abs/arXiv:2001.00096} {arXiv:2001.00096} \BibitemShut
  {NoStop}%
\bibitem [{\citenamefont {Andelkovic}\ \emph {et~al.}(2020)\citenamefont
  {Andelkovic}, \citenamefont {Milovanovic}, \citenamefont {Covaci},\ and\
  \citenamefont {Peeters}}]{Andelkovic2020}%
  \BibitemOpen
  \bibfield  {author} {\bibinfo {author} {\bibfnamefont {M.}~\bibnamefont
  {Andelkovic}}, \bibinfo {author} {\bibfnamefont {S.~P.}\ \bibnamefont
  {Milovanovic}}, \bibinfo {author} {\bibfnamefont {L.}~\bibnamefont {Covaci}},
  \ and\ \bibinfo {author} {\bibfnamefont {F.~M.}\ \bibnamefont {Peeters}},\
  }\href@noop {} {\bibfield  {journal} {\bibinfo  {journal} {Nano Letters}\
  }\textbf {\bibinfo {volume} {20}},\ \bibinfo {pages} {979} (\bibinfo {year}
  {2020})}\BibitemShut {NoStop}%
\bibitem [{\citenamefont {Zhu}\ \emph {et~al.}(2020{\natexlab{c}})\citenamefont
  {Zhu}, \citenamefont {Cazeaux}, \citenamefont {Luskin},\ and\ \citenamefont
  {Kaxiras}}]{ZiyanZhu2020_relax}%
  \BibitemOpen
  \bibfield  {author} {\bibinfo {author} {\bibfnamefont {Z.}~\bibnamefont
  {Zhu}}, \bibinfo {author} {\bibfnamefont {P.}~\bibnamefont {Cazeaux}},
  \bibinfo {author} {\bibfnamefont {M.}~\bibnamefont {Luskin}}, \ and\ \bibinfo
  {author} {\bibfnamefont {E.}~\bibnamefont {Kaxiras}},\ }\href@noop {}
  {\bibfield  {journal} {\bibinfo  {journal} {Phys. Rev. B}\ }\textbf {\bibinfo
  {volume} {101}},\ \bibinfo {pages} {224107} (\bibinfo {year}
  {2020}{\natexlab{c}})}\BibitemShut {NoStop}%
\bibitem [{\citenamefont {Tsai}\ \emph {et~al.}(2019)\citenamefont {Tsai},
  \citenamefont {Zhang}, \citenamefont {Zhu}, \citenamefont {Luo},
  \citenamefont {Carr}, \citenamefont {Luskin}, \citenamefont {Kaxiras},\ and\
  \citenamefont {Wang}}]{KanTing2019}%
  \BibitemOpen
  \bibfield  {author} {\bibinfo {author} {\bibfnamefont {K.-T.}\ \bibnamefont
  {Tsai}}, \bibinfo {author} {\bibfnamefont {X.}~\bibnamefont {Zhang}},
  \bibinfo {author} {\bibfnamefont {Z.}~\bibnamefont {Zhu}}, \bibinfo {author}
  {\bibfnamefont {Y.}~\bibnamefont {Luo}}, \bibinfo {author} {\bibfnamefont
  {S.}~\bibnamefont {Carr}}, \bibinfo {author} {\bibfnamefont {M.}~\bibnamefont
  {Luskin}}, \bibinfo {author} {\bibfnamefont {E.}~\bibnamefont {Kaxiras}}, \
  and\ \bibinfo {author} {\bibfnamefont {K.}~\bibnamefont {Wang}},\ }\href@noop
  {} {\enquote {\bibinfo {title} {Correlated superconducting and insulating
  states in twisted trilayer graphene moir\'e of moir\'e superlattices},}\ }
  (\bibinfo {year} {2019}),\ \Eprint {http://arxiv.org/abs/arXiv:1912.03375}
  {arXiv:1912.03375} \BibitemShut {NoStop}%
\bibitem [{\citenamefont {Trugman}(1983)}]{Trugman1983}%
  \BibitemOpen
  \bibfield  {author} {\bibinfo {author} {\bibfnamefont {S.~A.}\ \bibnamefont
  {Trugman}},\ }\href@noop {} {\bibfield  {journal} {\bibinfo  {journal} {Phys.
  Rev. B}\ }\textbf {\bibinfo {volume} {27}},\ \bibinfo {pages} {7539}
  (\bibinfo {year} {1983})}\BibitemShut {NoStop}%
\bibitem [{\citenamefont {Chalker}\ and\ \citenamefont
  {Coddington}(1988)}]{Chalker_1988}%
  \BibitemOpen
  \bibfield  {author} {\bibinfo {author} {\bibfnamefont {J.~T.}\ \bibnamefont
  {Chalker}}\ and\ \bibinfo {author} {\bibfnamefont {P.~D.}\ \bibnamefont
  {Coddington}},\ }\href@noop {} {\bibfield  {journal} {\bibinfo  {journal}
  {Journal of Physics C: Solid State Physics}\ }\textbf {\bibinfo {volume}
  {21}},\ \bibinfo {pages} {2665} (\bibinfo {year} {1988})}\BibitemShut
  {NoStop}%
\bibitem [{\citenamefont {Marston}\ and\ \citenamefont
  {Tsai}(1999)}]{Marston1999}%
  \BibitemOpen
  \bibfield  {author} {\bibinfo {author} {\bibfnamefont {J.~B.}\ \bibnamefont
  {Marston}}\ and\ \bibinfo {author} {\bibfnamefont {S.-W.}\ \bibnamefont
  {Tsai}},\ }\href@noop {} {\bibfield  {journal} {\bibinfo  {journal} {Phys.
  Rev. Lett.}\ }\textbf {\bibinfo {volume} {82}},\ \bibinfo {pages} {4906}
  (\bibinfo {year} {1999})}\BibitemShut {NoStop}%
\bibitem [{\citenamefont {Castro~Neto}\ \emph {et~al.}(2009)\citenamefont
  {Castro~Neto}, \citenamefont {Guinea}, \citenamefont {Peres}, \citenamefont
  {Novoselov},\ and\ \citenamefont {Geim}}]{Castro2009_grapheneRMP}%
  \BibitemOpen
  \bibfield  {author} {\bibinfo {author} {\bibfnamefont {A.~H.}\ \bibnamefont
  {Castro~Neto}}, \bibinfo {author} {\bibfnamefont {F.}~\bibnamefont {Guinea}},
  \bibinfo {author} {\bibfnamefont {N.~M.~R.}\ \bibnamefont {Peres}}, \bibinfo
  {author} {\bibfnamefont {K.~S.}\ \bibnamefont {Novoselov}}, \ and\ \bibinfo
  {author} {\bibfnamefont {A.~K.}\ \bibnamefont {Geim}},\ }\href@noop {}
  {\bibfield  {journal} {\bibinfo  {journal} {Rev. Mod. Phys.}\ }\textbf
  {\bibinfo {volume} {81}},\ \bibinfo {pages} {109} (\bibinfo {year}
  {2009})}\BibitemShut {NoStop}%
\bibitem [{\citenamefont {Liu}\ \emph {et~al.}(2003)\citenamefont {Liu},
  \citenamefont {Feng},\ and\ \citenamefont {Shen}}]{LiuLei2003_hBN}%
  \BibitemOpen
  \bibfield  {author} {\bibinfo {author} {\bibfnamefont {L.}~\bibnamefont
  {Liu}}, \bibinfo {author} {\bibfnamefont {Y.~P.}\ \bibnamefont {Feng}}, \
  and\ \bibinfo {author} {\bibfnamefont {Z.~X.}\ \bibnamefont {Shen}},\
  }\href@noop {} {\bibfield  {journal} {\bibinfo  {journal} {Phys. Rev. B}\
  }\textbf {\bibinfo {volume} {68}},\ \bibinfo {pages} {104102} (\bibinfo
  {year} {2003})}\BibitemShut {NoStop}%
\bibitem [{\citenamefont {Carr}\ \emph {et~al.}(2019)\citenamefont {Carr},
  \citenamefont {Fang}, \citenamefont {Zhu},\ and\ \citenamefont
  {Kaxiras}}]{Stephen2019}%
  \BibitemOpen
  \bibfield  {author} {\bibinfo {author} {\bibfnamefont {S.}~\bibnamefont
  {Carr}}, \bibinfo {author} {\bibfnamefont {S.}~\bibnamefont {Fang}}, \bibinfo
  {author} {\bibfnamefont {Z.}~\bibnamefont {Zhu}}, \ and\ \bibinfo {author}
  {\bibfnamefont {E.}~\bibnamefont {Kaxiras}},\ }\href {\doibase
  10.1103/PhysRevResearch.1.013001} {\bibfield  {journal} {\bibinfo  {journal}
  {Phys. Rev. Research}\ }\textbf {\bibinfo {volume} {1}},\ \bibinfo {pages}
  {013001} (\bibinfo {year} {2019})}\BibitemShut {NoStop}%
\bibitem [{\citenamefont {Kim}\ \emph {et~al.}(2018)\citenamefont {Kim},
  \citenamefont {Leconte}, \citenamefont {Chittari}, \citenamefont {Watanabe},
  \citenamefont {Taniguchi}, \citenamefont {MacDonald}, \citenamefont {Jung},\
  and\ \citenamefont {Jung}}]{Hakseong2018}%
  \BibitemOpen
  \bibfield  {author} {\bibinfo {author} {\bibfnamefont {H.}~\bibnamefont
  {Kim}}, \bibinfo {author} {\bibfnamefont {N.}~\bibnamefont {Leconte}},
  \bibinfo {author} {\bibfnamefont {B.~L.}\ \bibnamefont {Chittari}}, \bibinfo
  {author} {\bibfnamefont {K.}~\bibnamefont {Watanabe}}, \bibinfo {author}
  {\bibfnamefont {T.}~\bibnamefont {Taniguchi}}, \bibinfo {author}
  {\bibfnamefont {A.~H.}\ \bibnamefont {MacDonald}}, \bibinfo {author}
  {\bibfnamefont {J.}~\bibnamefont {Jung}}, \ and\ \bibinfo {author}
  {\bibfnamefont {S.}~\bibnamefont {Jung}},\ }\href {\doibase
  10.1021/acs.nanolett.8b03423} {\bibfield  {journal} {\bibinfo  {journal}
  {Nano Letters}\ }\textbf {\bibinfo {volume} {18}},\ \bibinfo {pages} {7732}
  (\bibinfo {year} {2018})}\BibitemShut {NoStop}%
\bibitem [{\citenamefont {Hunt}\ \emph {et~al.}(2013)\citenamefont {Hunt},
  \citenamefont {Sanchez-Yamagishi}, \citenamefont {Young}, \citenamefont
  {Yankowitz}, \citenamefont {LeRoy}, \citenamefont {Watanabe}, \citenamefont
  {Taniguchi}, \citenamefont {Moon}, \citenamefont {Koshino}, \citenamefont
  {Jarillo-Herrero},\ and\ \citenamefont {Ashoori}}]{Hunt2013}%
  \BibitemOpen
  \bibfield  {author} {\bibinfo {author} {\bibfnamefont {B.}~\bibnamefont
  {Hunt}}, \bibinfo {author} {\bibfnamefont {J.~D.}\ \bibnamefont
  {Sanchez-Yamagishi}}, \bibinfo {author} {\bibfnamefont {A.~F.}\ \bibnamefont
  {Young}}, \bibinfo {author} {\bibfnamefont {M.}~\bibnamefont {Yankowitz}},
  \bibinfo {author} {\bibfnamefont {B.~J.}\ \bibnamefont {LeRoy}}, \bibinfo
  {author} {\bibfnamefont {K.}~\bibnamefont {Watanabe}}, \bibinfo {author}
  {\bibfnamefont {T.}~\bibnamefont {Taniguchi}}, \bibinfo {author}
  {\bibfnamefont {P.}~\bibnamefont {Moon}}, \bibinfo {author} {\bibfnamefont
  {M.}~\bibnamefont {Koshino}}, \bibinfo {author} {\bibfnamefont
  {P.}~\bibnamefont {Jarillo-Herrero}}, \ and\ \bibinfo {author} {\bibfnamefont
  {R.~C.}\ \bibnamefont {Ashoori}},\ }\href {\doibase 10.1126/science.1237240}
  {\bibfield  {journal} {\bibinfo  {journal} {Science}\ }\textbf {\bibinfo
  {volume} {340}},\ \bibinfo {pages} {1427} (\bibinfo {year}
  {2013})}\BibitemShut {NoStop}%
\bibitem [{\citenamefont {Song}\ \emph {et~al.}(2013)\citenamefont {Song},
  \citenamefont {Shytov},\ and\ \citenamefont {Levitov}}]{Justin2013}%
  \BibitemOpen
  \bibfield  {author} {\bibinfo {author} {\bibfnamefont {J.~C.~W.}\
  \bibnamefont {Song}}, \bibinfo {author} {\bibfnamefont {A.~V.}\ \bibnamefont
  {Shytov}}, \ and\ \bibinfo {author} {\bibfnamefont {L.~S.}\ \bibnamefont
  {Levitov}},\ }\href@noop {} {\bibfield  {journal} {\bibinfo  {journal} {Phys.
  Rev. Lett.}\ }\textbf {\bibinfo {volume} {111}},\ \bibinfo {pages} {266801}
  (\bibinfo {year} {2013})}\BibitemShut {NoStop}%
\bibitem [{\citenamefont {Fukui}\ \emph {et~al.}(2005)\citenamefont {Fukui},
  \citenamefont {Hatsugai},\ and\ \citenamefont {Suzuki}}]{Takahiro2005}%
  \BibitemOpen
  \bibfield  {author} {\bibinfo {author} {\bibfnamefont {T.}~\bibnamefont
  {Fukui}}, \bibinfo {author} {\bibfnamefont {Y.}~\bibnamefont {Hatsugai}}, \
  and\ \bibinfo {author} {\bibfnamefont {H.}~\bibnamefont {Suzuki}},\
  }\href@noop {} {\bibfield  {journal} {\bibinfo  {journal} {Journal of the
  Physical Society of Japan}\ }\textbf {\bibinfo {volume} {74}},\ \bibinfo
  {pages} {1674} (\bibinfo {year} {2005})}\BibitemShut {NoStop}%
\bibitem [{\citenamefont {Nomura}\ and\ \citenamefont
  {MacDonald}(2006)}]{Nomura2006}%
  \BibitemOpen
  \bibfield  {author} {\bibinfo {author} {\bibfnamefont {K.}~\bibnamefont
  {Nomura}}\ and\ \bibinfo {author} {\bibfnamefont {A.~H.}\ \bibnamefont
  {MacDonald}},\ }\href@noop {} {\bibfield  {journal} {\bibinfo  {journal}
  {Phys. Rev. Lett.}\ }\textbf {\bibinfo {volume} {96}},\ \bibinfo {pages}
  {256602} (\bibinfo {year} {2006})}\BibitemShut {NoStop}%
\bibitem [{\citenamefont {B\"uttiker}(1986)}]{Buttiker1986}%
  \BibitemOpen
  \bibfield  {author} {\bibinfo {author} {\bibfnamefont {M.}~\bibnamefont
  {B\"uttiker}},\ }\href {\doibase 10.1103/PhysRevLett.57.1761} {\bibfield
  {journal} {\bibinfo  {journal} {Phys. Rev. Lett.}\ }\textbf {\bibinfo
  {volume} {57}},\ \bibinfo {pages} {1761} (\bibinfo {year}
  {1986})}\BibitemShut {NoStop}%
\bibitem [{\citenamefont {Wang}\ \emph {et~al.}(2013)\citenamefont {Wang},
  \citenamefont {Lian}, \citenamefont {Zhang},\ and\ \citenamefont
  {Zhang}}]{JingWang2013}%
  \BibitemOpen
  \bibfield  {author} {\bibinfo {author} {\bibfnamefont {J.}~\bibnamefont
  {Wang}}, \bibinfo {author} {\bibfnamefont {B.}~\bibnamefont {Lian}}, \bibinfo
  {author} {\bibfnamefont {H.}~\bibnamefont {Zhang}}, \ and\ \bibinfo {author}
  {\bibfnamefont {S.-C.}\ \bibnamefont {Zhang}},\ }\href@noop {} {\bibfield
  {journal} {\bibinfo  {journal} {Phys. Rev. Lett.}\ }\textbf {\bibinfo
  {volume} {111}},\ \bibinfo {pages} {086803} (\bibinfo {year}
  {2013})}\BibitemShut {NoStop}%
\bibitem [{\citenamefont {Jackiw}\ and\ \citenamefont
  {Rebbi}(1976)}]{Jackiw1976}%
  \BibitemOpen
  \bibfield  {author} {\bibinfo {author} {\bibfnamefont {R.}~\bibnamefont
  {Jackiw}}\ and\ \bibinfo {author} {\bibfnamefont {C.}~\bibnamefont {Rebbi}},\
  }\href@noop {} {\bibfield  {journal} {\bibinfo  {journal} {Phys. Rev. Lett.}\
  }\textbf {\bibinfo {volume} {36}},\ \bibinfo {pages} {1116} (\bibinfo {year}
  {1976})}\BibitemShut {NoStop}%
\bibitem [{\citenamefont {Su}\ \emph {et~al.}(1979)\citenamefont {Su},
  \citenamefont {Schrieffer},\ and\ \citenamefont {Heeger}}]{SuSchrieffer1979}%
  \BibitemOpen
  \bibfield  {author} {\bibinfo {author} {\bibfnamefont {W.~P.}\ \bibnamefont
  {Su}}, \bibinfo {author} {\bibfnamefont {J.~R.}\ \bibnamefont {Schrieffer}},
  \ and\ \bibinfo {author} {\bibfnamefont {A.~J.}\ \bibnamefont {Heeger}},\
  }\href@noop {} {\bibfield  {journal} {\bibinfo  {journal} {Phys. Rev. Lett.}\
  }\textbf {\bibinfo {volume} {42}},\ \bibinfo {pages} {1698} (\bibinfo {year}
  {1979})}\BibitemShut {NoStop}%
\bibitem [{\citenamefont {Kim}\ \emph {et~al.}(2016)\citenamefont {Kim},
  \citenamefont {Yankowitz}, \citenamefont {Fallahazad}, \citenamefont {Kang},
  \citenamefont {Movva}, \citenamefont {Huang}, \citenamefont {Larentis},
  \citenamefont {Corbet}, \citenamefont {Taniguchi}, \citenamefont {Watanabe},
  \citenamefont {Banerjee}, \citenamefont {LeRoy},\ and\ \citenamefont
  {Tutuc}}]{Kim2016_twisttech}%
  \BibitemOpen
  \bibfield  {author} {\bibinfo {author} {\bibfnamefont {K.}~\bibnamefont
  {Kim}}, \bibinfo {author} {\bibfnamefont {M.}~\bibnamefont {Yankowitz}},
  \bibinfo {author} {\bibfnamefont {B.}~\bibnamefont {Fallahazad}}, \bibinfo
  {author} {\bibfnamefont {S.}~\bibnamefont {Kang}}, \bibinfo {author}
  {\bibfnamefont {H.~C.~P.}\ \bibnamefont {Movva}}, \bibinfo {author}
  {\bibfnamefont {S.}~\bibnamefont {Huang}}, \bibinfo {author} {\bibfnamefont
  {S.}~\bibnamefont {Larentis}}, \bibinfo {author} {\bibfnamefont {C.~M.}\
  \bibnamefont {Corbet}}, \bibinfo {author} {\bibfnamefont {T.}~\bibnamefont
  {Taniguchi}}, \bibinfo {author} {\bibfnamefont {K.}~\bibnamefont {Watanabe}},
  \bibinfo {author} {\bibfnamefont {S.~K.}\ \bibnamefont {Banerjee}}, \bibinfo
  {author} {\bibfnamefont {B.~J.}\ \bibnamefont {LeRoy}}, \ and\ \bibinfo
  {author} {\bibfnamefont {E.}~\bibnamefont {Tutuc}},\ }\href {\doibase
  10.1021/acs.nanolett.5b05263} {\bibfield  {journal} {\bibinfo  {journal}
  {Nano Letters}\ }\textbf {\bibinfo {volume} {16}},\ \bibinfo {pages} {1989}
  (\bibinfo {year} {2016})},\ \bibinfo {note} {pMID: 26859527}\BibitemShut
  {NoStop}%
\bibitem [{\citenamefont {Woods}\ \emph {et~al.}(2014)\citenamefont {Woods},
  \citenamefont {Britnell}, \citenamefont {Eckmann}, \citenamefont {Ma},
  \citenamefont {Lu}, \citenamefont {Guo}, \citenamefont {Lin}, \citenamefont
  {Yu}, \citenamefont {Cao}, \citenamefont {Gorbachev}, \citenamefont
  {Kretinin}, \citenamefont {Park}, \citenamefont {Ponomarenko}, \citenamefont
  {Katsnelson}, \citenamefont {Gornostyrev}, \citenamefont {Watanabe},
  \citenamefont {Taniguchi}, \citenamefont {Casiraghi}, \citenamefont {Gao},
  \citenamefont {Geim},\ and\ \citenamefont {Novoselov}}]{Woods2014}%
  \BibitemOpen
  \bibfield  {author} {\bibinfo {author} {\bibfnamefont {C.~R.}\ \bibnamefont
  {Woods}}, \bibinfo {author} {\bibfnamefont {L.}~\bibnamefont {Britnell}},
  \bibinfo {author} {\bibfnamefont {A.}~\bibnamefont {Eckmann}}, \bibinfo
  {author} {\bibfnamefont {R.~S.}\ \bibnamefont {Ma}}, \bibinfo {author}
  {\bibfnamefont {J.~C.}\ \bibnamefont {Lu}}, \bibinfo {author} {\bibfnamefont
  {H.~M.}\ \bibnamefont {Guo}}, \bibinfo {author} {\bibfnamefont
  {X.}~\bibnamefont {Lin}}, \bibinfo {author} {\bibfnamefont {G.~L.}\
  \bibnamefont {Yu}}, \bibinfo {author} {\bibfnamefont {Y.}~\bibnamefont
  {Cao}}, \bibinfo {author} {\bibfnamefont {R.~V.}\ \bibnamefont {Gorbachev}},
  \bibinfo {author} {\bibfnamefont {A.~V.}\ \bibnamefont {Kretinin}}, \bibinfo
  {author} {\bibfnamefont {J.}~\bibnamefont {Park}}, \bibinfo {author}
  {\bibfnamefont {L.~A.}\ \bibnamefont {Ponomarenko}}, \bibinfo {author}
  {\bibfnamefont {M.~I.}\ \bibnamefont {Katsnelson}}, \bibinfo {author}
  {\bibfnamefont {Y.~N.}\ \bibnamefont {Gornostyrev}}, \bibinfo {author}
  {\bibfnamefont {K.}~\bibnamefont {Watanabe}}, \bibinfo {author}
  {\bibfnamefont {T.}~\bibnamefont {Taniguchi}}, \bibinfo {author}
  {\bibfnamefont {C.}~\bibnamefont {Casiraghi}}, \bibinfo {author}
  {\bibfnamefont {H.-J.}\ \bibnamefont {Gao}}, \bibinfo {author} {\bibfnamefont
  {A.~K.}\ \bibnamefont {Geim}}, \ and\ \bibinfo {author} {\bibfnamefont
  {K.~S.}\ \bibnamefont {Novoselov}},\ }\href@noop {} {\bibfield  {journal}
  {\bibinfo  {journal} {Nature Physics}\ }\textbf {\bibinfo {volume} {10}},\
  \bibinfo {pages} {451} (\bibinfo {year} {2014})}\BibitemShut {NoStop}%
\bibitem [{\citenamefont {Kazmierczak}\ \emph {et~al.}(2020)\citenamefont
  {Kazmierczak}, \citenamefont {Winkle}, \citenamefont {Ophus}, \citenamefont
  {Bustillo}, \citenamefont {Brown}, \citenamefont {Carr}, \citenamefont
  {Ciston}, \citenamefont {Taniguchi}, \citenamefont {Watanabe},\ and\
  \citenamefont {Bediako}}]{Nathanael2020_strain}%
  \BibitemOpen
  \bibfield  {author} {\bibinfo {author} {\bibfnamefont {N.~P.}\ \bibnamefont
  {Kazmierczak}}, \bibinfo {author} {\bibfnamefont {M.~V.}\ \bibnamefont
  {Winkle}}, \bibinfo {author} {\bibfnamefont {C.}~\bibnamefont {Ophus}},
  \bibinfo {author} {\bibfnamefont {K.~C.}\ \bibnamefont {Bustillo}}, \bibinfo
  {author} {\bibfnamefont {H.~G.}\ \bibnamefont {Brown}}, \bibinfo {author}
  {\bibfnamefont {S.}~\bibnamefont {Carr}}, \bibinfo {author} {\bibfnamefont
  {J.}~\bibnamefont {Ciston}}, \bibinfo {author} {\bibfnamefont
  {T.}~\bibnamefont {Taniguchi}}, \bibinfo {author} {\bibfnamefont
  {K.}~\bibnamefont {Watanabe}}, \ and\ \bibinfo {author} {\bibfnamefont
  {D.~K.}\ \bibnamefont {Bediako}},\ }\href@noop {} {\enquote {\bibinfo {title}
  {Strain fields in twisted bilayer graphene},}\ } (\bibinfo {year} {2020}),\
  \Eprint {http://arxiv.org/abs/arXiv:2008.09761} {arXiv:2008.09761}
  \BibitemShut {NoStop}%
\bibitem [{\citenamefont {Bostwick}\ \emph {et~al.}(2012)\citenamefont
  {Bostwick}, \citenamefont {Rotenberg}, \citenamefont {Avila},\ and\
  \citenamefont {Asensio}}]{Aaron2012_nanoARPES}%
  \BibitemOpen
  \bibfield  {author} {\bibinfo {author} {\bibfnamefont {A.}~\bibnamefont
  {Bostwick}}, \bibinfo {author} {\bibfnamefont {E.}~\bibnamefont {Rotenberg}},
  \bibinfo {author} {\bibfnamefont {J.}~\bibnamefont {Avila}}, \ and\ \bibinfo
  {author} {\bibfnamefont {M.~C.}\ \bibnamefont {Asensio}},\ }\href@noop {}
  {\bibfield  {journal} {\bibinfo  {journal} {Synchrotron Radiation News}\
  }\textbf {\bibinfo {volume} {25}},\ \bibinfo {pages} {19} (\bibinfo {year}
  {2012})}\BibitemShut {NoStop}%
\bibitem [{\citenamefont {Dudin}\ \emph {et~al.}(2010)\citenamefont {Dudin},
  \citenamefont {Lacovig}, \citenamefont {Fava}, \citenamefont {Nicolini},
  \citenamefont {Bianco}, \citenamefont {Cautero},\ and\ \citenamefont
  {Barinov}}]{Dudin2010_nanoARPES}%
  \BibitemOpen
  \bibfield  {author} {\bibinfo {author} {\bibfnamefont {P.}~\bibnamefont
  {Dudin}}, \bibinfo {author} {\bibfnamefont {P.}~\bibnamefont {Lacovig}},
  \bibinfo {author} {\bibfnamefont {C.}~\bibnamefont {Fava}}, \bibinfo {author}
  {\bibfnamefont {E.}~\bibnamefont {Nicolini}}, \bibinfo {author}
  {\bibfnamefont {A.}~\bibnamefont {Bianco}}, \bibinfo {author} {\bibfnamefont
  {G.}~\bibnamefont {Cautero}}, \ and\ \bibinfo {author} {\bibfnamefont
  {A.}~\bibnamefont {Barinov}},\ }\href@noop {} {\bibfield  {journal} {\bibinfo
   {journal} {Journal of Synchrotron Radiation}\ }\textbf {\bibinfo {volume}
  {17}},\ \bibinfo {pages} {445} (\bibinfo {year} {2010})}\BibitemShut
  {NoStop}%
\bibitem [{\citenamefont {Zhu}\ \emph {et~al.}(2020{\natexlab{d}})\citenamefont
  {Zhu}, \citenamefont {Shi},\ and\ \citenamefont
  {MacDonald}}]{Jihang2020_ARPES}%
  \BibitemOpen
  \bibfield  {author} {\bibinfo {author} {\bibfnamefont {J.}~\bibnamefont
  {Zhu}}, \bibinfo {author} {\bibfnamefont {J.}~\bibnamefont {Shi}}, \ and\
  \bibinfo {author} {\bibfnamefont {A.~H.}\ \bibnamefont {MacDonald}},\
  }\href@noop {} {\enquote {\bibinfo {title} {Theory of arpes in graphene-based
  moir\'e superlattices},}\ } (\bibinfo {year} {2020}{\natexlab{d}}),\ \Eprint
  {http://arxiv.org/abs/arXiv:2006.08908} {arXiv:2006.08908} \BibitemShut
  {NoStop}%
\bibitem [{\citenamefont {Wehling}\ \emph {et~al.}(2015)\citenamefont
  {Wehling}, \citenamefont {Huber}, \citenamefont {Lichtenstein},\ and\
  \citenamefont {Katsnelson}}]{Wehling2015}%
  \BibitemOpen
  \bibfield  {author} {\bibinfo {author} {\bibfnamefont {T.~O.}\ \bibnamefont
  {Wehling}}, \bibinfo {author} {\bibfnamefont {A.}~\bibnamefont {Huber}},
  \bibinfo {author} {\bibfnamefont {A.~I.}\ \bibnamefont {Lichtenstein}}, \
  and\ \bibinfo {author} {\bibfnamefont {M.~I.}\ \bibnamefont {Katsnelson}},\
  }\href@noop {} {\bibfield  {journal} {\bibinfo  {journal} {Phys. Rev. B}\
  }\textbf {\bibinfo {volume} {91}},\ \bibinfo {pages} {041404} (\bibinfo
  {year} {2015})}\BibitemShut {NoStop}%
\bibitem [{\citenamefont {Hipolito}\ and\ \citenamefont
  {Pereira}(2017)}]{Hipolito_2017}%
  \BibitemOpen
  \bibfield  {author} {\bibinfo {author} {\bibfnamefont {F.}~\bibnamefont
  {Hipolito}}\ and\ \bibinfo {author} {\bibfnamefont {V.~M.}\ \bibnamefont
  {Pereira}},\ }\href@noop {} {\bibfield  {journal} {\bibinfo  {journal} {2D
  Materials}\ }\textbf {\bibinfo {volume} {4}},\ \bibinfo {pages} {021027}
  (\bibinfo {year} {2017})}\BibitemShut {NoStop}%
\bibitem [{\citenamefont {Mak}\ \emph {et~al.}(2018)\citenamefont {Mak},
  \citenamefont {Xiao},\ and\ \citenamefont {Shan}}]{Mak2018}%
  \BibitemOpen
  \bibfield  {author} {\bibinfo {author} {\bibfnamefont {K.~F.}\ \bibnamefont
  {Mak}}, \bibinfo {author} {\bibfnamefont {D.}~\bibnamefont {Xiao}}, \ and\
  \bibinfo {author} {\bibfnamefont {J.}~\bibnamefont {Shan}},\ }\href@noop {}
  {\bibfield  {journal} {\bibinfo  {journal} {Nature Photonics}\ }\textbf
  {\bibinfo {volume} {12}},\ \bibinfo {pages} {451} (\bibinfo {year}
  {2018})}\BibitemShut {NoStop}%
\bibitem [{\citenamefont {Wang}\ \emph
  {et~al.}(2019{\natexlab{b}})\citenamefont {Wang}, \citenamefont {Zihlmann},
  \citenamefont {Liu}, \citenamefont {Makk}, \citenamefont {Watanabe},
  \citenamefont {Taniguchi}, \citenamefont {Baumgartner},\ and\ \citenamefont
  {Sch\"{o}nenberger}}]{WangLujun2019}%
  \BibitemOpen
  \bibfield  {author} {\bibinfo {author} {\bibfnamefont {L.}~\bibnamefont
  {Wang}}, \bibinfo {author} {\bibfnamefont {S.}~\bibnamefont {Zihlmann}},
  \bibinfo {author} {\bibfnamefont {M.-H.}\ \bibnamefont {Liu}}, \bibinfo
  {author} {\bibfnamefont {P.}~\bibnamefont {Makk}}, \bibinfo {author}
  {\bibfnamefont {K.}~\bibnamefont {Watanabe}}, \bibinfo {author}
  {\bibfnamefont {T.}~\bibnamefont {Taniguchi}}, \bibinfo {author}
  {\bibfnamefont {A.}~\bibnamefont {Baumgartner}}, \ and\ \bibinfo {author}
  {\bibfnamefont {C.}~\bibnamefont {Sch\"{o}nenberger}},\ }\href {\doibase
  10.1021/acs.nanolett.8b05061} {\bibfield  {journal} {\bibinfo  {journal}
  {Nano Letters}\ }\textbf {\bibinfo {volume} {19}},\ \bibinfo {pages} {2371}
  (\bibinfo {year} {2019}{\natexlab{b}})},\ \bibinfo {note} {pMID:
  30803238}\BibitemShut {NoStop}%
\bibitem [{\citenamefont {Finney}\ \emph {et~al.}(2019)\citenamefont {Finney},
  \citenamefont {Yankowitz}, \citenamefont {Muraleetharan}, \citenamefont
  {Watanabe}, \citenamefont {Taniguchi}, \citenamefont {Dean},\ and\
  \citenamefont {Hone}}]{Finney2019}%
  \BibitemOpen
  \bibfield  {author} {\bibinfo {author} {\bibfnamefont {N.~R.}\ \bibnamefont
  {Finney}}, \bibinfo {author} {\bibfnamefont {M.}~\bibnamefont {Yankowitz}},
  \bibinfo {author} {\bibfnamefont {L.}~\bibnamefont {Muraleetharan}}, \bibinfo
  {author} {\bibfnamefont {K.}~\bibnamefont {Watanabe}}, \bibinfo {author}
  {\bibfnamefont {T.}~\bibnamefont {Taniguchi}}, \bibinfo {author}
  {\bibfnamefont {C.~R.}\ \bibnamefont {Dean}}, \ and\ \bibinfo {author}
  {\bibfnamefont {J.}~\bibnamefont {Hone}},\ }\href@noop {} {\bibfield
  {journal} {\bibinfo  {journal} {Nature Nanotechnology}\ }\textbf {\bibinfo
  {volume} {14}},\ \bibinfo {pages} {1029} (\bibinfo {year}
  {2019})}\BibitemShut {NoStop}%
\bibitem [{\citenamefont {Amorim}\ and\ \citenamefont
  {Castro}(2018)}]{Amorim2018}%
  \BibitemOpen
  \bibfield  {author} {\bibinfo {author} {\bibfnamefont {B.}~\bibnamefont
  {Amorim}}\ and\ \bibinfo {author} {\bibfnamefont {E.~V.}\ \bibnamefont
  {Castro}},\ }\href@noop {} {\enquote {\bibinfo {title} {Electronic spectral
  properties of incommensurate twisted trilayer graphene},}\ } (\bibinfo {year}
  {2018}),\ \Eprint {http://arxiv.org/abs/arXiv:1807.11909} {arXiv:1807.11909}
  \BibitemShut {NoStop}%
\bibitem [{\citenamefont {Lin}\ \emph {et~al.}(2020)\citenamefont {Lin},
  \citenamefont {Su},\ and\ \citenamefont {Ni}}]{Lin2020_instability}%
  \BibitemOpen
  \bibfield  {author} {\bibinfo {author} {\bibfnamefont {X.}~\bibnamefont
  {Lin}}, \bibinfo {author} {\bibfnamefont {K.}~\bibnamefont {Su}}, \ and\
  \bibinfo {author} {\bibfnamefont {J.}~\bibnamefont {Ni}},\ }\href@noop {}
  {\enquote {\bibinfo {title} {Misalignment instability in magic-angle twisted
  bilayer graphene on hexagonal boron nitride},}\ } (\bibinfo {year} {2020}),\
  \Eprint {http://arxiv.org/abs/arXiv:2011.01541} {arXiv:2011.01541}
  \BibitemShut {NoStop}%
\bibitem [{\citenamefont {Mao}\ and\ \citenamefont
  {Senthil}(2020)}]{DanMao2020}%
  \BibitemOpen
  \bibfield  {author} {\bibinfo {author} {\bibfnamefont {D.}~\bibnamefont
  {Mao}}\ and\ \bibinfo {author} {\bibfnamefont {T.}~\bibnamefont {Senthil}},\
  }\href@noop {} {\enquote {\bibinfo {title} {Quasiperiodicity, band topology,
  and moiré graphene},}\ } (\bibinfo {year} {2020}),\ \Eprint
  {http://arxiv.org/abs/arXiv:2011.06034} {arXiv:2011.06034} \BibitemShut
  {NoStop}%
\end{thebibliography}%

\end{document}